\documentclass[jgr]{agutex}

\usepackage{multirow,color}
\usepackage{bbding}
\usepackage{amssymb}
\usepackage{ulem}
\usepackage{verbatim}
\usepackage{url}

\usepackage{graphicx}
\usepackage[displaymath]{lineno}

\def\gsim{\;\lower4pt\hbox{${\buildrel\displaystyle >\over\sim}$}\;}
\def\lsim{\;\lower4pt\hbox{${\buildrel\displaystyle <\over\sim}$}\;}
\def\grls{\;\lower4pt\hbox{${\buildrel\displaystyle >\over <}$}\;}

\newcommand{\ve}[1]{\mathbf{#1}}
\paperid{2017JA024971}\journalid{2018}\articleid{}{}
\authorrunninghead{WANG ET AL.}
\titlerunninghead{Twist Distribution in an Interplanetary MC}

\begin{document}

\title{Understanding the twist distribution inside magnetic flux ropes by anatomizing an interplanetary magnetic cloud}

\author{Yuming Wang,$^{1,2,*}$ Chenglong Shen,$^{1,2}$ Rui Liu,$^{1,3}$ Jiajia Liu,$^4$ Jingnan Guo,$^5$ Xiaolei Li,$^{1,3}$ Mengjiao Xu,$^{1,3}$ Qiang Hu,$^{6}$ Tielong Zhang,$^{1,3,7}$}

\affil{$^1$ CAS Key Laboratory of Geospace Environment, School of Earth and Space Sciences, University of Science and Technology of China, Hefei 230026, China}

\affil{$^2$ Synergetic Innovation Center of Quantum Information \& Quantum Physics, University of Science and Technology of China, Hefei 230026, China}

\affil{$^3$ Collaborative Innovation Center of Astronautical Science and Technology, Hefei 230026, China}

\affil{$^4$ Solar Physics and Space Plasma Research Center, School of Mathematics and Statistics, University of Sheffield, Sheffield S37RH, UK}

\affil{$^5$ Institute of Experimental and Applied Physics, University of Kiel, Germany}

\affil{$^6$ Department of Space Science and CSPAR, The University of Alabama in Huntsville, Huntsville, Alabama, USA}

\affil{$^7$ Space Research Institute, Austrian Academy of Sciences, Graz, Austria}

\affil{$^*$ Corresponding author, Email: ymwang@ustc.edu.cn}

\begin{abstract}
Magnetic flux rope (MFR) is the core
structure of the greatest eruptions, i.e., the coronal mass ejections (CMEs), on the Sun, and magnetic clouds are 
post-eruption MFRs in interplanetary space. There is a strong debate about whether or not a MFR exists prior to a CME 
and how the MFR forms/grows through magnetic reconnection during the eruption. 
%Twist of magnetic field lines in a MFR contains the key information, but is not well understood yet. 
Here we report a rare event, in which a magnetic cloud was observed sequentially by four
spacecraft near Mercury, Venus, Earth and Mars, respectively. With the aids of a uniform-twist flux rope model
and a newly developed method that can recover a shock-compressed structure, 
we find that the axial magnetic flux and helicity of the magnetic cloud decreased when it propagated 
outward but the twist increased. 
Our analysis suggests that the `pancaking' effect and `erosion'
effect may jointly cause such variations. The significance of the `pancaking' effect  
is difficult to be estimated, but the signature of the erosion can be found as the 
imbalance of the azimuthal flux of the cloud. The latter implies that 
the magnetic cloud was eroded significantly leaving its inner core exposed to the solar wind at far distance. The 
increase of the twist together with the presence of the erosion effect suggests that 
the post-eruption MFR may have a high-twist core enveloped 
by a less-twisted outer shell.
These results pose a great challenge to the current understanding on the solar eruptions as well as the formation and instability of MFRs.
\end{abstract}

\begin{article}

\section{Introduction}

Magnetic flux rope (MFR) is a fundamental plasma structure in the universe, and tightly related to various eruptive
phenomena due to non-potential field therein. It could appear in magnetic reconnection regions manifesting as 
	magnetic islands~\citep[e.g.,][]{Daughton_etal_2011}, in the corona and heliosphere known as coronal mass ejections (CMEs) and 
	magnetic clouds~\citep[e.g.,][]{Zhang_etal_2012, Burlaga_etal_1981, Vourlidas_etal_2013}, 
	and in astrophysical jets with the scale up to thousands of light years~\citep[e.g.,][]{Owen_etal_1989, Marscher_etal_2008}. 
	Previous theoretical studies~\citep[e.g.,][]{Kruskal_etal_1958, Shafranov_1963, 
Hood_Priest_1981} suggested that a MFR will be subject to kink instability  
once the total twist angle, $\Phi$, of its magnetic field lines exceeds a certain threshold, e.g., $2.5\pi$ radians for  
flux ropes in the solar atmosphere~\citep{Hood_Priest_1981} with the confirmation by laboratory experiments~\citep{Myers_etal_2015}. 
	This threshold, however, is challenged by frequent observations of high-twist flux ropes not only in the solar atmosphere~\citep[e.g.,][]{Vrsnak_etal_1991, Gary_Moore_2004, Srivastava_etal_2010} 
	but also in the heliosphere~\citep[e.g.][]{Hu_etal_2015, Wang_etal_2016} and even in galaxies~\citep[e.g.,][]{Owen_etal_1989, Marscher_etal_2008, Perley_etal_1984, Gomez_etal_2008}. 
%\mod{[*** The theoretical models and numerical simulations on astrophysical jets suggest that high twist is 
%possible\cite{Uchida_Shibata_1985, Nakamura_etal_2001, LiH_etal_2001, Koide_etal_2002, Lynden-Bell_etal_2006}.]}
The most recent statistical study of 115 
	interplanetary magnetic clouds near the Earth~\citep{Wang_etal_2016} showed that the total twist angle can be more than $10\pi$ radians, much larger than the above theoretical thresholds, and its upper limit follows
the relation given by \citet{Dungey_Loughhead_1954}: 
\begin{eqnarray}
\Phi_c=2\frac{l}{R} \label{eq:threshold}
\end{eqnarray} 
where $l$ is the length of the MFR's axis and $R$ is the radius of the MFR's cross-section. Although a uniform-twist force-free flux rope model was used in \citet{Wang_etal_2016}'s study, the relation does suggest
that a thinner and/or longer MFR can have higher-twisted magnetic field lines, or the inner core of a MFR can be more twisted, and does imply that
a very long MFR, such as those in astrophysical jets, may be kink stable, even though it is highly-twisted. 

However, in light of the magnetohydrodynamic theory, a linear force-free
flux rope stays at a lower state of magnetic energy than a nonlinear force-free or non-force-free flux rope with the same helicity. Thus, interplanetary
magnetic clouds, which are considered to be the post-eruption MFRs having relaxed for a sufficient period of time, were usually modeled 
as a linear force-free flux rope following Lundquist solution~\citep{Lundquist_1950, Lepping_etal_2006}, suggesting a minimum twist at the axis 
of the MFR and a maximum twist at the periphery. This is opposite to the
implication from equation (\ref{eq:threshold}) 
that the inner core of a MFR can have a higher twist.
This inconsistency raises the question of how the 
twist distributes in the cross-section of a naturally hatched MFR, e.g., those in CMEs, and is closely related
to the long-standing debate whether or not a MFR forms prior to CME eruptions.

There are two competing scenarios about the onset of CMEs in terms of MFRs. One suggests that CMEs do not need a preexisting MFR, which can newly develop
from sheared arcades through converging motion and magnetic reconnection during the course of the eruption~\citep[e.g.,][]{Antiochos_etal_1999, Moore_etal_2001, Karpen_etal_2012}. 
The other believes that there must be a seed MFR, no matter how small it is, before the eruption~\citep[e.g.,][]{Kopp_Pneuman_1976, Titov_Demoulin_1999}. 
The consensus is that the magnetic reconnection taking place beneath the erupting MFR will add 
a considerable amount of magnetic fluxes into the MFR by converting overlying field lines to the outer shell of the MFR~\citep[e.g.,][]{Qiu_etal_2007}.
If the seed MFR in the second scenario formed in a way similar to that in the first scenario 
through the magnetic reconnection of inner sheared arcades, the post-eruption MFRs of two scenarios might not be distinguishable~\citep[e.g.,][]{Aulanier_etal_2010}. 
However, there are at least two other ways to generate a MFR in the solar atmosphere. One is the rotational/shearing motion of fluid elements on the photosphere which are frozen 
into a bunch of closed magnetic field lines, and the other is the emergence of a MFR from the convention zone beneath the photosphere.
Thus, the two scenarios may make the post-eruption MFR quite different in terms of the distribution of twist. In the former case, the twist should increase 
from the axis to periphery of the MFR as illustrated by the cartoon in the paper by \citet{Moore_etal_2001}. In the latter case, the twist 
might have a stage-like distribution in the cross-section of the MFR, consisting of a core MFR and a outer shell with a different twist.
Here another debate is  whether the field lines added through reconnection are highly 
twisted~\citep{Longcope_Beveridge_2007, Aulanier_etal_2012} or weakly twisted~\citep{vanBallegooijen_Martens_1989}. 

In this paper, we present a rare event, in which an interplanetary magnetic cloud was sequentially observed
by four spacecraft near the inner planets: Mercury, Venus, Earth and Mars. By anatomizing the magnetic properties
of the magnetic cloud at different heliocentric distance, we try to address the aforementioned debates, and 
refine the global picture of interplanetary magnetic clouds erupted from the Sun.

\section{Overview of the event}

The cases of a magnetic cloud observed in-situ by multiple spacecraft at different heliocentric distances were occasionally reported 
	in the past 40 years~\citep[e.g.,][]{Burlaga_etal_1981, Mulligan_etal_2001, Leitner_etal_2007, Du_etal_2007, Nakwacki_etal_2011, 
Nieves-Chinchilla_etal_2012, Good_etal_2015, Winslow_etal_2016}. 
The most famous one is the first identified magnetic cloud observed by Helios 1 and 2, IMP 8 and Voyager 1 and 2 in the inner and out heliosphere 
in 1978 January~\citep{Burlaga_etal_1981}. However, that event is not suitable for our study, because the data are too poor.
To our knowledge, there is a small number of well-observed events 
due to limited number of spacecraft in the heliosphere at same time, 
which were/are not necessarily well aligned along the radial direction. 

	\subsection{The magnetic cloud at Mercury}
The magnetic cloud in this study was first observed by the magnetometer onboard spacecraft MErcury Surface, Space ENvironment, GEochemistry and Ranging 
	(MESSENGER, \citealt{Anderson_etal_2007}) orbiting around Mercury. Figure~\ref{fig:mc_mer_ven}a shows the 
measurements of the magnetic field during 2014 February 15 -- 16. Since Mercury owns a significant intrinsic magnetic field, it has a magnetosphere
and a bow shock upstream~\citep{Slavin_2004}, and MESSENGER was immersed in pure solar wind for a limited time in its each $\sim8$-hr 
orbit. The regions within the magnetosheath and magnetosphere can be identified by the crossings of the bow shock,
characterized by a sudden change in the magnetic field strength, as indicated by the dark-shadowed regions in the figure.
The front boundaries of the shadowed regions are the crossings close to the nose of the bow shock, and the rear boundaries 
locate at the flank. There are several spikes in the magnetic field strength during February 15 20:00 UT -- 16 01:00 UT, which were 
probably due to the swings of the bow shock disturbed by the passage of the magnetic cloud.

The magnetic cloud can be recognized between February 15 20:20 UT and about 15:40 UT on the next day as indicated by the light-shadowed region 
bounded by two vertical blue lines in Figure~\ref{fig:mc_mer_ven}a. Without those dark-shadowed regions, 
the signatures of a typical magnetic cloud are evident, including
enhanced magnetic field strength (up to more than $45$ nT compared with the field less than $25$ nT before the cloud) 
and the large and smooth rotation of the field vector. Unfortunately, there are only sporadic measurements
of solar wind plasma, and therefore we do not include them here. 
%It is noteworthy that this event was missed in the CME list\cite{Winslow_etal_2015} compiled based on the observations of MESSENGER. 
Despite of some small data gaps due to the 
passages of Mercury's magnetosheath and magnetosphere, neither a strong driven shock which is typically accompanied by a narrow and highly 
fluctuated shock sheath, nor a wide shock sheath which usually follows a weak shock, could be found outside of either end of the cloud.
Thus, the cloud should travel with a speed comparable to the ambient solar wind, 
consistent with the nearly symmetric profile of the magnetic field shown in the first panel of Figure~\ref{fig:mc_mer_ven}a.

	\subsection{The magnetic cloud at Earth}
\begin{figure*}[ht]
\begin{center}
\includegraphics[width=\hsize]{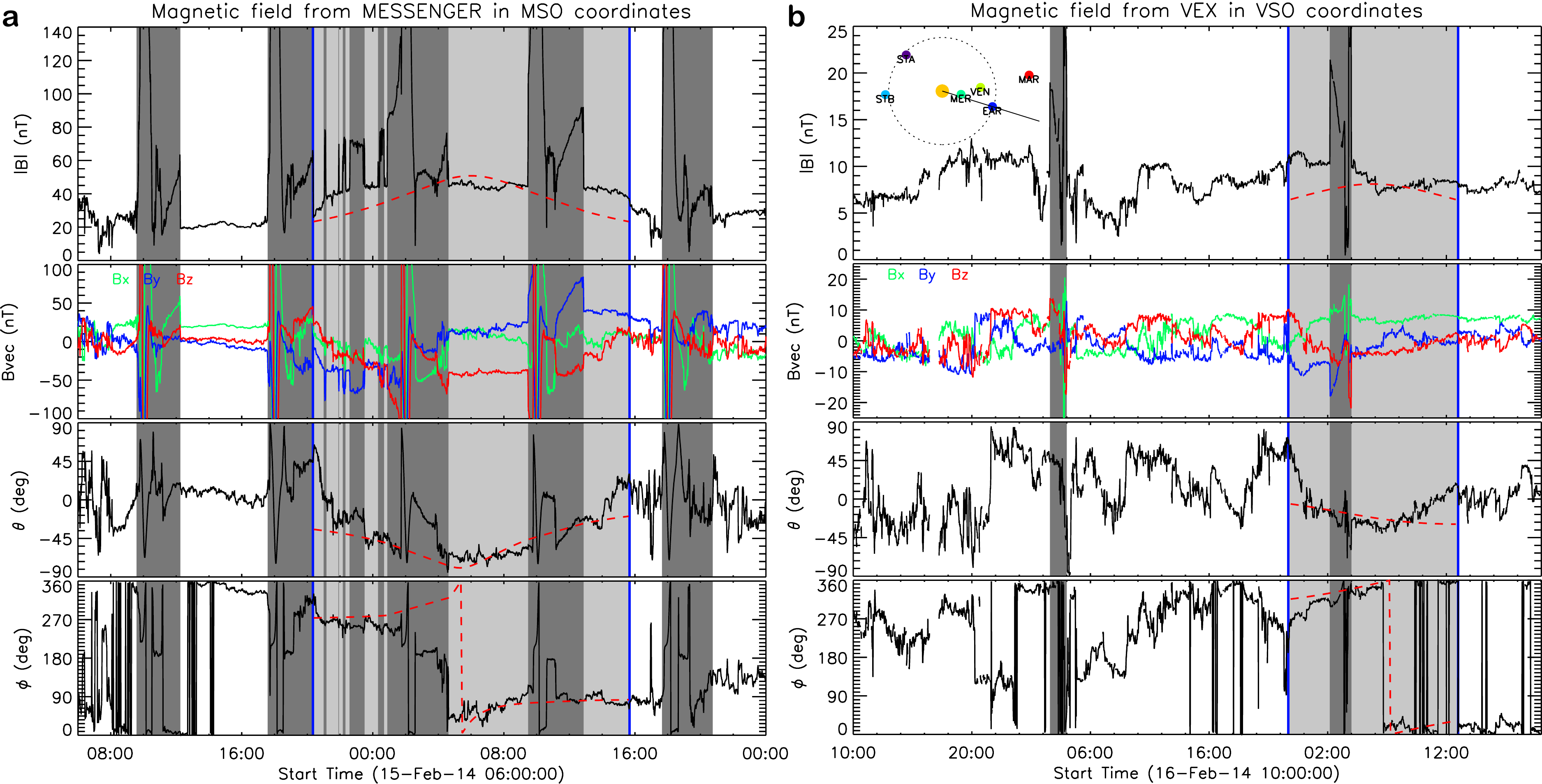}
	\caption{Magnetic fields measured by MESSENGER (Panel {\bf a}) and VEX (Panel {\bf b}). In each panel, from the top to bottom,
	there are the total magnetic field strength, $|B|$, three components of $\ve B$ in the planet-solar-orbital coordinate 
	system, i.e., MSO and VSO coordinates for MESSENGER and VEX data, respectively, and the elevation, $\theta$, and 
	azimuthal, $\phi$, angle of the $\ve B$ vector. The dark-shadowed regions indicate the magnetosheath and magnetosphere
	behind the planetary bow shock. The magnetic cloud of interest is in the light-shadowed region bounded by two vertical 
	blue lines. The red dashed lines are the fitting results by the velocity-modified uniform-twist force-free flux rope 
	model (see Sec.\ref{sec:fitting}). The inset on the upper-left corner of Panel b shows the positions of the planets and spacecraft.
	}\label{fig:mc_mer_ven}
\end{center}
\end{figure*}

\begin{figure*}[t]
\begin{center}
\includegraphics[width=0.8\hsize]{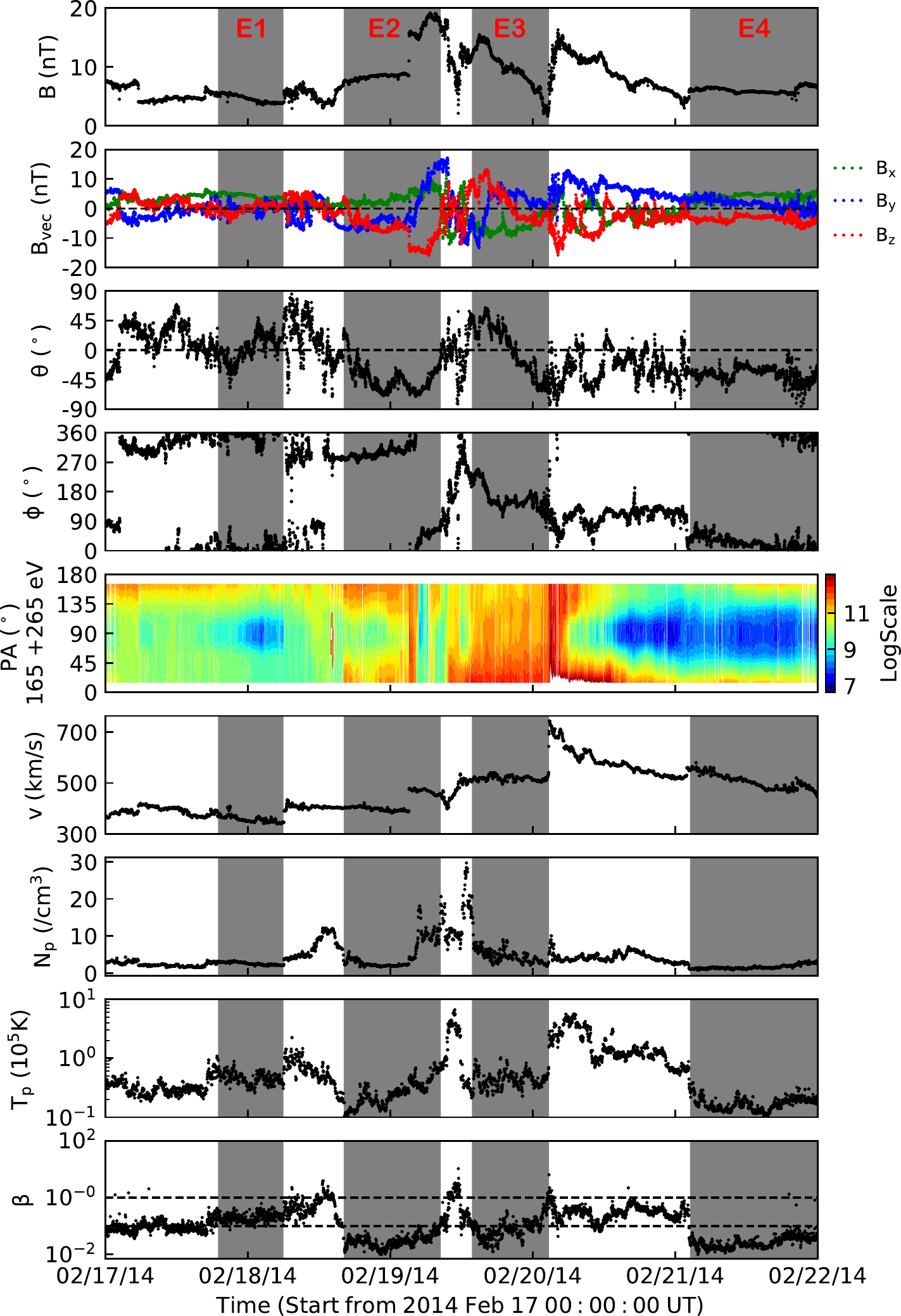}
	\caption{Observations of the magnetic cloud recorded by the Wind spacecraft at the Earth. 
	From the top to bottom, it shows the total magnetic field strength, $|B|$, the three components of $\ve B$
	in GSE coordinates, the elevation, $\theta$, and azimuthal, $\phi$, angle of the $\ve B$ vector, the pitch angle (PA) distribution
	of suprathermal electrons, the bulk velocity, $v$, of the solar wind, and the number density, $N_p$, temperature, $T_p$, and
	plasma $\beta$ of protons. The four shadow regions indicate four ejecta, labeled as `E1' through `E4'. 
	}\label{fig:mc_earth}
\end{center}
\end{figure*}

Before identifying the counterpart of the magnetic cloud at Venus, we check its signature at the Earth because Earth was well-aligned
with the Sun and Mercury (about $2.6^\circ$ apart away from Mercury) at that time (see the inset at the upper-left corner of Fig.\ref{fig:mc_mer_ven}b or Fig.\ref{fig:dips}) 
and the Wind spacecraft~\citep{Lepping_etal_1995, Ogilvie_etal_1995} near the Earth has complete 
sets of the interplanetary magnetic field and solar wind plasma data. Figure~\ref{fig:mc_earth} shows the measurements in 5 days from February 17 
to 21. Combining the signatures of a CME ejecta, such as enhanced magnetic field strength, smooth rotation of field vector, low temperature,
low proton $\beta$ and bi-directional suprathermal electron beams, etc., we may find four ejecta marked by `E1' through `E4'
in the shadowed regions. Ejecta `E1' arrived at the Earth at about 19:00 UT on February 17. Considering the distance between the Earth and 
Mercury is about $0.65$ AU, we can estimate that the transit time of `E1' is about $46.7$ hrs corresponding to a transit speed of about $580$ km s$^{-1}$,
which is much higher than its in-situ speed of about $370$ km s$^{-1}$. If `E1' was the counterpart of the magnetic cloud 
observed by MESSENGER, it must have experienced a great deceleration, and can be estimated to have a speed about $800$ km s$^{-1}$ near Mercury.
Such a fast ejecta should drive a strong shock as well as a shock sheath, which was not observed. Thus, `E1' cannot be the counterpart of the magnetic cloud.
The same  analysis on ejecta `E3' and `E4' suggests that their expected transit speeds are $300$ and $210$ km s$^{-1}$, respectively, 
much lower than the in-situ speeds, both faster than 500 km s$^{-1}$. Thus, the two ejecta are also not the counterpart of the magnetic cloud.
As to ejecta `E2' arriving at about 16:10 UT on February 18, the expected transit speed is about $400$ km s$^{-1}$,
well consistent with the in-situ speed measured by the Wind. Thus, it should be the same magnetic cloud observed at Mercury, unambiguously.
The association can be further confirmed, as the counterparts of ejecta
`E1', `E3' and `E4' at Mercury as well as their corresponding CMEs can all be identified (we put the detailed identification process in
Appendix~\ref{supp:corres_cmes} to make the main text fluent).

Ejecta `E2' has clear signatures of a magnetic cloud. The rotation of the magnetic field was evident and smooth, the pitch angle of the suprathermal
electrons concentrated around $0^\circ$ and $180^\circ$, and the proton $\beta$ was lower than $0.1$. The rear part of the magnetic cloud 
was compressed by a strong forward shock, driven by ejecta `E3', which destroyed the signature of the bi-directional electron beams. 

	\subsection{The magnetic cloud at Venus}\label{sec:mc_venus}
Venus locates between the Earth and Mercury at $0.72$ AU. It was not well-aligned with the two planets during the period of interest, but in a close angular position.
The angular separation of Venus and Mercury at the times of the magnetic cloud passing through them was about $25^\circ$.
We may expect to observe the same magnetic cloud between February 16 and 18. 
Similar to the situation at Mercury, there are only scattered measurements of pure solar wind plasma by Venus EXpress (VEX, \citealt{Svedhem_etal_2007}),
and sometimes VEX located behind the bow shock and in the Venus induced magnetosphere. The magnetic field data from the VEX 
magnetometer~\citep{ZhangT_etal_2006} suggests that the interval between February 17 22:40 UT and 18 13:00 UT is the only possible candidate, during
which the long and smooth rotation of magnetic field vector is evident (see Fig.~\ref{fig:mc_mer_ven}b). 

\begin{figure}[b]
\begin{center}
\includegraphics[width=\hsize]{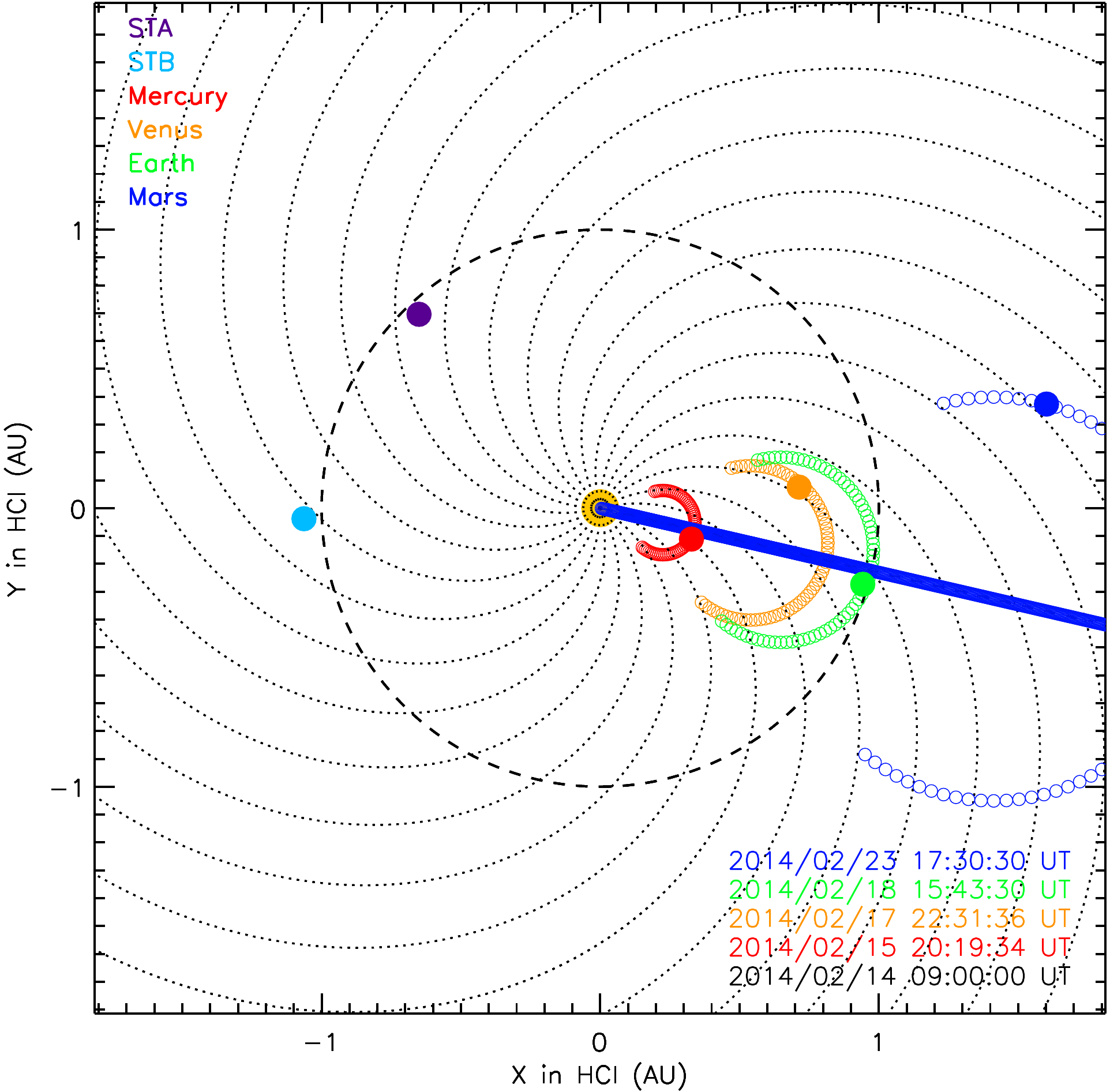}
	\caption{Trajectory of the magnetic cloud front on the ecliptic plane in the heliocentric inertial coordinates (HCI) 
	estimated by the DIPS model (see Sec.\ref{sec:mc_venus}). The red, orange, green, and blue dots mark the positions of Mercury, Venus,
	Earth and Mars, respectively, at the times of the magnetic cloud encountering them which have been given at the lower-right corner. 
The purple and azure dots indicate the positions of the Solar Terrestrial Relations Observatories (STEREO) A and B, respectively. 
	The magnetic cloud front on the ecliptic plane is modeled by a circular arc with its two ends 
	tangential to two radial directions between which the angle indicates the angular width of the magnetic cloud.}\label{fig:dips}
\end{center}
\end{figure}

One may question that, if this structure is the same magnetic cloud, why the magnetic cloud spent about $50$ hrs to travel from Mercury to 
Venus (a distance of $\sim0.37$ AU) but less than only 18 hrs from Venus to the Earth (a distance of $\sim0.28$ AU). 
This is likely due to the curved front of the magnetic cloud~\citep{Mostl_Davies_2013, 
Shen_etal_2014}. By assuming a certain propagation speed of the magnetic cloud,
we may model the arrival times of the magnetic cloud at different distances as shown in 
Figure~\ref{fig:dips}. 
The model used in the study was developed for the CME Deflection
in InterPlanetary Space (called DIPS model)~\citep{Wang_etal_2004b, Wang_etal_2016a, ZhuangB_etal_2017}. The input parameters
include the propagation speed, angular width and initial propagation direction of the CME and the speed
of background solar wind. In this study, we set the speeds of both the magnetic cloud and solar wind 
constant as $400$ km s$^{-1}$ because
the transit speed of the magnetic cloud from Mercury to the Earth is $400$ km s$^{-1}$, very close to the background
solar wind speed measured by Wind, indicating little momentum exchange between the cloud and solar wind. The angular width
and initial propagation direction are adjusted to obtain the best matched case, which are found to be about $60^\circ$
and right facing to the Earth. The circular arcs in Figure~\ref{fig:dips} approximate the front of the magnetic cloud,
of which the two ends are tangent to two $60^\circ$-separated lines, respectively, starting from the Sun.
Please note that the arcs just model the front of the cross-section of the magnetic cloud cut by
the ecliptic plane, but not the front of the global magnetic cloud structure, as implied by the modeled orientation of the cloud (see Table~\ref{tb:fitting}).
It is revealed that the observed arrivals at Mercury, Venus and the Earth can be well matched when the magnetic cloud 
propagated along the Sun-Mercury-Earth line and the angular width is about $60^\circ$.

\subsection{Extrapolating the trajectory of the magnetic cloud to Mars and back to the Sun}
\begin{figure}[b]
\begin{center}
\includegraphics[width=\hsize]{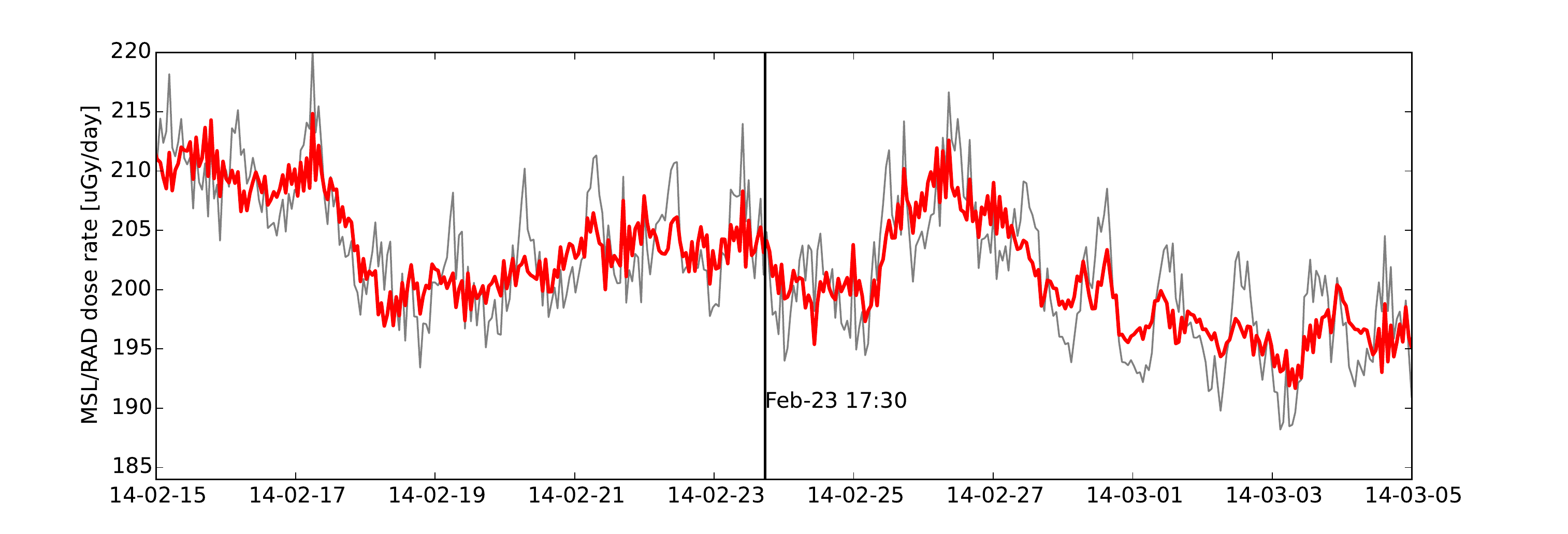}
	\caption{The dose rate of the cosmic rays observed by RAD onboard the MSL at Mars. 
	The gray line marks the original data which shows a daily periodic variation caused by the diurnal Martian 
	atmospheric thermal tide~\citep{GuoJ_etal_2017}. The red line represents the data when applied a frequency filter to remove 
	the diurnal variations therein~\citep{GuoJ_etal_2018}. The vertical line marks the predicted arrival of the magnetic cloud at Mars.}\label{fig:mc_mars}
\end{center}
\end{figure}

Mars locates at about $1.65$ AU around that time.
By extrapolating the trajectory of the cloud to the orbit of Mars, we may predict that
the arrival of the magnetic cloud at Mars was about 17:30 UT on February 23, when it was only $7^\circ$ apart away from 
Venus or $32^\circ$ from Mercury by comparing their positions at the times of the cloud crossing them. 
Unfortunately, there was no appropriate instrument
measuring the interplanetary magnetic field or the solar wind plasma near Mars. The only useful data are from the Radiation 
Assessment Detector (RAD, \citealt{Hassler_etal_2012}) onboard Mars Science Laboratory (MSL, \citealt{Grotzinger_etal_2012}), 
providing the information of 
Forbush decreases which are believed to be caused by the passage of CMEs~\citep{Cane_2000}. Figure~\ref{fig:mc_mars} shows 
the dose rate of cosmic rays recorded by the RAD from February 15 to March 5, during which several Forbush decreases are evident.
The predicted arrival of the magnetic cloud perfectly corresponds to the beginning of a decrease as marked by the vertical line. 
According to the Wind observations, there were several faster ejecta catching up with the magnetic cloud of interest. 
Thus, it is very possible that these ejecta interacted with each other and formed a complex structure before arriving at 
Mars to make such a significant Forbush decrease. 

Similarly, we may extrapolate the trajectory of the magnetic cloud back to the Sun. 
The predicted onset time of the corresponding CME is 09:00 UT on February 14. However, the magnetic cloud was a slow and therefore weak one, 
whereas the Sun was quite active around that period from February 13 to 15, during which many larger and stronger CMEs were launched. 
Thus, the identification of the CME corresponding to the magnetic cloud in coronagraphs is more or less ambiguous, and no definite eruptive
signature on the solar surface can be found around the expected time, suggesting the possibility of a stealth CME. The detailed process of our identification is given 
in Appendix~\ref{supp:mc_source}.

\section{Magnetic evolution of the magnetic cloud from Mercury to Earth}

\subsection{Reconstruct the magnetic cloud with the uniform-twist force-free flux rope model}\label{sec:fitting}
The observations of the same magnetic cloud at different heliocentric distance provide us a unique opportunity to study the magnetic 
properties of the cloud and their changes with the distance. 
%To reconstruct the magnetic cloud and estimate its magnetic properties, we use the nonlinear
%force-free flux rope model with the uniform-twist solution~\citep{Wang_etal_2016} to fit the 
%one-dimensional measurements along the path of the spacecraft (see Methods). 
There are various techniques to reconstruct a magnetic cloud from one-dimensional measurements along
the observational path~\citep[e.g.,][]{Burlaga_etal_1981, Goldstein_1983, Marubashi_1986, Lepping_etal_1990,
Mulligan_Russell_2001, Hu_Sonnerup_2002, Hidalgo_etal_2002, Cid_etal_2002, Vandas_Romashets_2003, Dasso_etal_2006, Wang_etal_2015, Wang_etal_2016}.
Cylindrical force-free flux rope models are frequently used, and tested to be reliable~\citep{Riley_etal_2004}.
Here, we choose the velocity-modified uniform-twist force-free flux rope model~\citep{Wang_etal_2016} to fit the observed data,
which treats the magnetic twist as a free parameter in the fitting procedure.
The Grad-Shafranov (GS) reconstruction technique~\citep{Hu_Sonnerup_2002} can also obtain the twist of a magnetic cloud,
but it needs more solar wind plasma parameters, including the total gas pressure, and therefore cannot be applied
to the MESSENGER and VEX data.
%Thus, to make consistent fittings to the magnetic cloud at different distances, we do not try to use the GS technique.

\begin{figure*}[t]
\begin{center}
\includegraphics[width=\hsize]{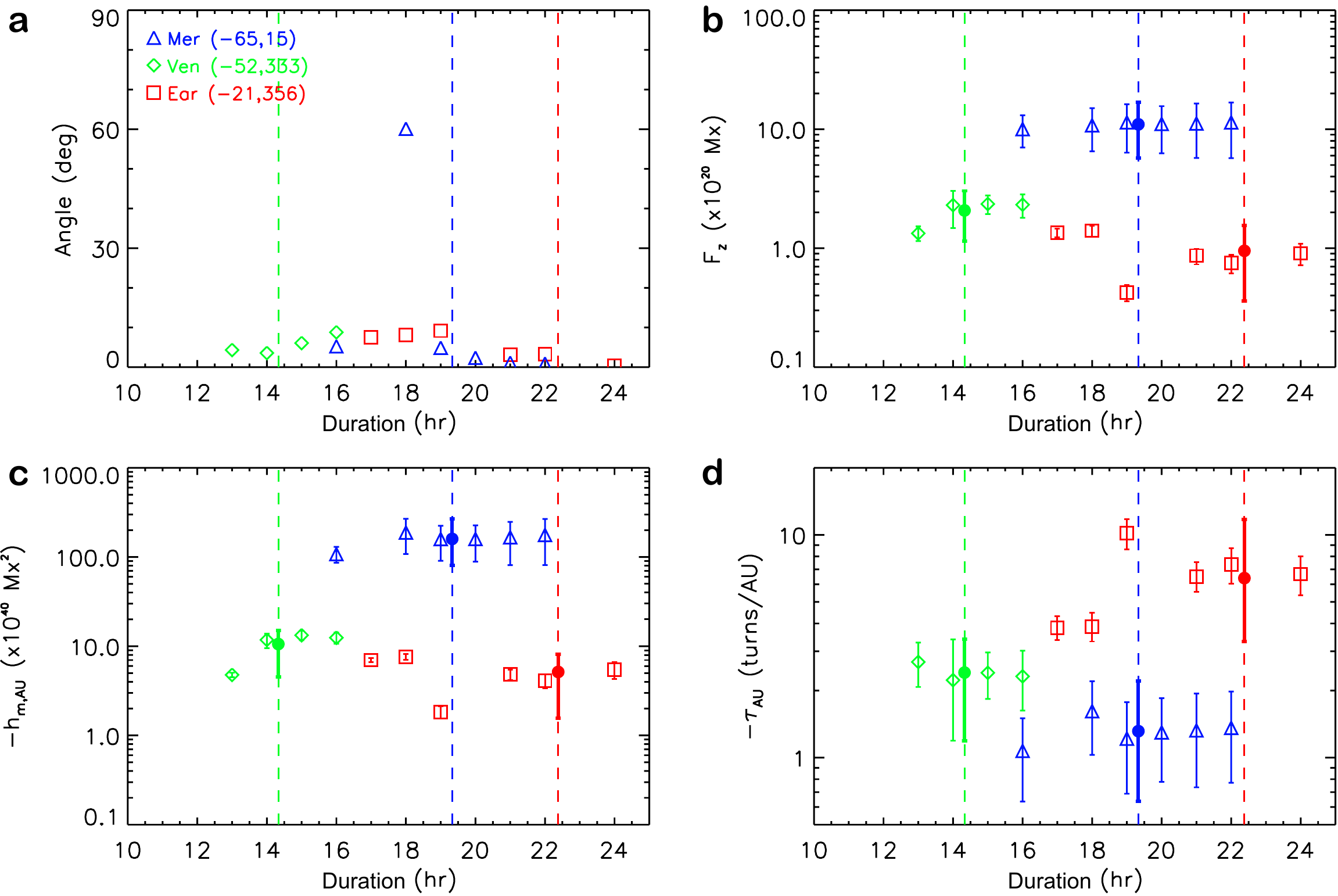}
	\caption{Fitting parameters of the magnetic cloud as a function of the boundary (see Sec.\ref{sec:fitting}). The horizontal axis indicates
	the duration from the front boundary to the rear boundary of the magnetic cloud, and the vertical dashed lines denote the duration
based on the identified 
	boundaries of the cloud based on the observations with the blue, green and red colors (or triangle, diamond and square symbols) 
	for the cloud at Mercury, Venus and the Earth, respectively. Each symbol stands for a test fitting with a pair of different front and rear boundaries. 
	Panel {\bf a}: The angle between the orientation (defined by $\theta$ and $\phi$ in the planet-solar-orbital coordinates) of the 
	axis of the magnetic cloud from each test fitting and the overall averaged orientation of the axis, which is listed in the upper-left corner. 
	Panel {\bf b}--{\bf d}: The values of the axial magnetic flux, $F_z$, the magnetic helicity per unit length, $h_{m,_{AU}}$, 
	and the number of twist per unit length, $\tau_{_{AU}}$. The subscript `AU' means that the parameters are rescaled to the values at the distance
of $1$ AU (see Sec.\ref{sec:res} for more details). The solid dots mark the average of the symbols with the same color, 
	and the error bars of the dots cover the uncertainties of the symbols. 
	%The horizontal dashed line at $\tau_{_{AU}}=-2.5$ turns per
	%AU is plotted to guide eyes, showing that the twist of the magnetic cloud at the Earth is significantly stronger than that at Mercury. 
	}\label{fig:fitting}
\end{center}
\end{figure*}

The fitting model we used has 10 free parameters: the magnetic field strength at the flux rope's axis ($B_0$), the orientation of the axis (the elevation and azimuthal angles, $\theta$ and $\phi$, 
in GSE coordinates), the closest approach of the observational path to the axis ($d$), 
three components of the propagation velocity ($v_x, v_y, v_z$), the expansion speed ($v_{exp}$) and poloidal speed ($v_{pol}$) at the boundary of the flux rope, and importantly, the twist.
These free parameters are coupled, and we constrain them with both the measurements of magnetic field and
solar wind velocity. Although there is no data of solar wind velocity from the MESSENGER and VEX,
we may assume that the magnetic cloud propagated at a constant speed
of $400$ km s$^{-1}$ without expansion, which is reasonable based on the above DIPS model result and the
flattened profile of the radial velocity recorded by Wind at 1 AU. The influence of the non-expansion assumption 
on the fitting results is tested for the magnetic cloud at Mercury by setting an expansion speed of about $20$ km s$^{-1}$, which is small (see Appendix~\ref{supp:expansion}). 
The time resolution of the data input to our model is set
to 5-min. The detailed description of the fitting
technique of this model can be found in our recent paper~\citep{Wang_etal_2016}.

As indicated by the name of the model, the twist is assumed to be uniform in the cross-section of a flux rope. This
is a good approximation to most magnetic clouds. In \citet{Hu_etal_2015}, it was shown
that the twist is probably high near the axis of a MFR and then quickly drops to a lower value when moving away from the axis, 
which suggested 
that the twist is almost uniform in most part of a MFR except the place very close to its axis. The observational work about
a solar eruption by \citet{WangW_etal_2017} reached a similar conclusion. Even if a magnetic cloud carries an irregular twist profile,  
our model will give a kind of averaged twist over the shell of the cloud crossed by the spacecraft, which could be treated as a first-order approximation.
If the spacecraft at the different planets crossed the cloud with different impact distances to its axis, we may anatomize how the 
twist distributes in the cloud.

\begin{figure*}
\begin{center}
\includegraphics[width=0.8\hsize]{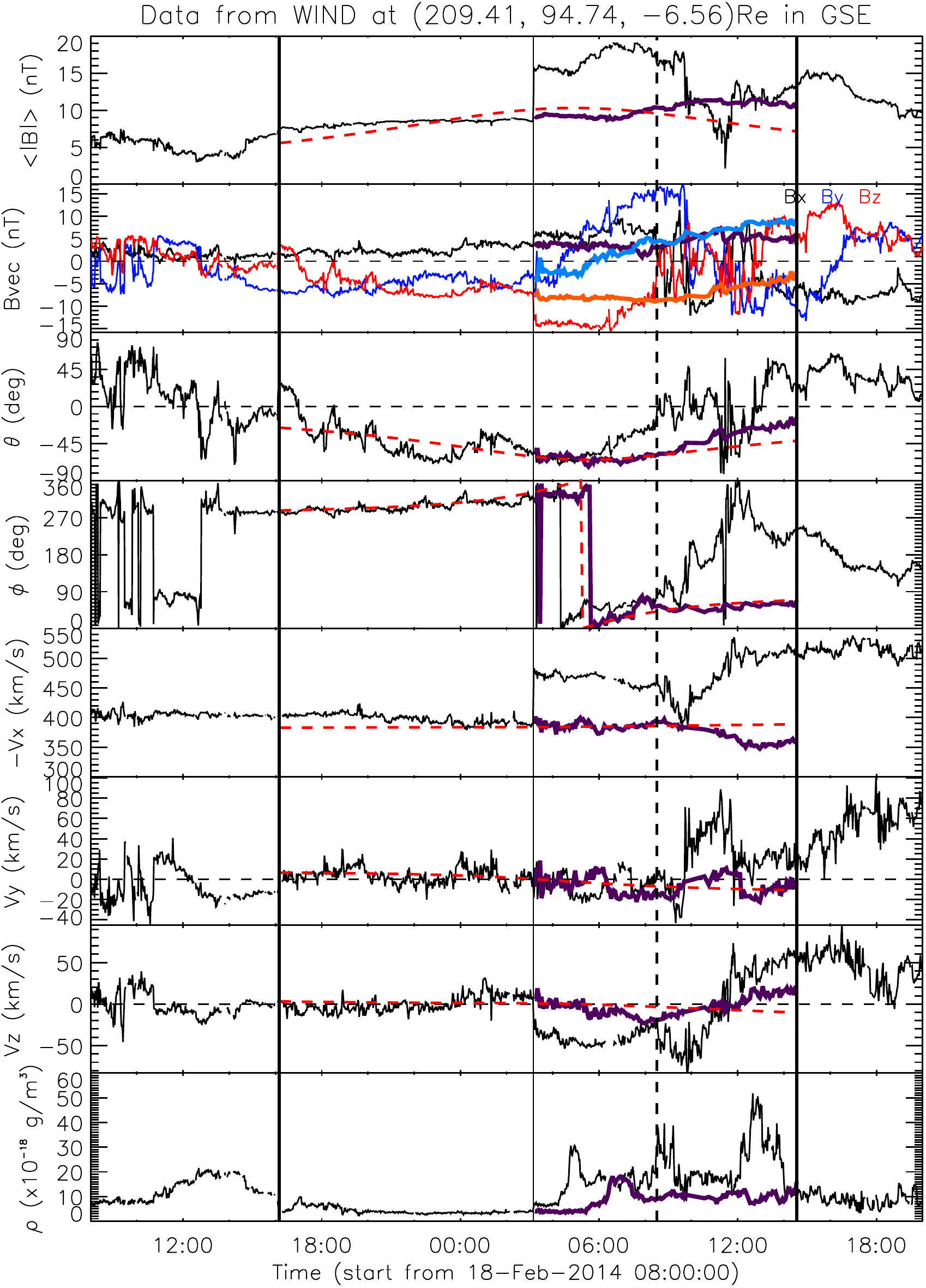}
	\caption{Zoom-in plot for ejecta `E2' observed at Earth. The first and last vertical lines mark the front and rear boundaries of 
the recovered magnetic cloud, and the vertical dashed line indicates the original rear boundary of the cloud. The second vertical line marks the shock surface.
The thick curves between the second and last vertical lines represent 
		the recovered structure, which originally locates between the second and third vertical lines (see Sec.\ref{sec:shocked}). The duration of the recovered structure is determined by equation (\ref{eq:dt}). The red dashed curves are the fitting results.
	}\label{fig:shock_recover}
\end{center}
\end{figure*}

The most important free parameter in the fitting is the orientation
of the magnetic cloud's axis, which can affect the reliability of other fitting parameters. One major factor influencing the orientation
is the location of the boundary of the cloud, which is difficult to be precisely determined.
Thus, to test the reliability of the fitting, we run test fittings by moving the front and rear boundaries 
simultaneously inward or outward with the same interval, and get a set of test fitting results. 
Not all of the fittings are successful. The quality of a fitting result can be assessed by the combination of
the normalized root-mean-square ($\chi_n$) of the difference between the modeled and observed data and a set
of three quantities related to the twist: the percentage (per) of the data points falling in the uncertainty range of the
modeled twist, the correlation coefficient (cc) of the modeled and measured twists and the confidence level
(cl) of the correlation (see Sec.2.2 on Page 9324 of \citealt{Wang_etal_2016} for more details). In this study, we set a criterion of
$\chi_n\leq 0.6$, per~$\geq 0.4$, cc~$\geq 0.4$ and cl~$\geq 0.9$ for an acceptable fitting. Figure~\ref{fig:fitting} shows the test fitting 
results, in which only the fittings satisfying the criterion are displayed. According to these successful fittings, the orientations
of the magnetic cloud axis derived based on different boundaries (see Fig.\ref{fig:fitting}a) concentrate to a certain value with differences less
than about $10^\circ$ (with only one exception for the fitting to the MESSENGER data, which is omitted in determining the final orientation below), 
suggesting a high reliability of the fitting result. The final orientation of the magnetic cloud axis at each distance is 
the average of the orientations of the successful test fittings (as listed in Table~\ref{tb:fitting}). The red dashed lines in Figure~\ref{fig:mc_mer_ven} and Figure~\ref{fig:shock_recover}
are the fitting curves obtained based on the final orientations.

It should be noted that the magnetic cloud observed at the Earth, which was partially    
compressed by an overtaking shock, cannot be fitted directly. We recover the shocked structure before applying the fitting technique by assuming
that the parameters in the shock sheath still follow the shock relation. Though it is a very ideal approximation, the fitting
results seem to be reliable. The next subsection gives the details.

\subsection{Recover the shocked structure}\label{sec:shocked}
The shock arrived at Wind at 03:10 UT on February 19, and the shock sheath spanned over about $6.5$ hrs, of which the first $5.3$-hr interval located inside
the magnetic cloud. To apply a fitting technique to the magnetic cloud, the shocked part of the magnetic
cloud has to be recovered back to the uncompressed state. To accomplish this purpose, we assume that (1) the
magnetic field, plasma velocity and density in the sheath region can be related to the uncompressed state by
the shock relation, i.e., Rankine–-Hugoniot jump conditions, and (2) the shock normal, $\hat{\ve n}$, shock speed, $v_s$,
and the compression ratio, $r_c$, are the same as those at the observed shock surface.
Treating the sheath region as the downstream
(using subscript `$2$') of the shock, the uncompressed state, i.e., the parameters of the upstream (using subscript `$1$')
of the shock, can be given by
\begin{eqnarray}
\left\{\begin{array}{l}
\rho_1=\frac{1}{r_c}\rho_2\\
\ve B_{1n}=\ve B_{2n}\\
\ve B_{1\perp}=\frac{v_{A2}^2-u_2^2}{v_{A2}^2-r_cu_2^2}\ve B_{2\perp} \\
\ve u_{1n}=r_c\ve u_{2n}\\
\ve u_{1\perp}=\frac{v_{A2}^2-u_2^2}{v_{A2}^2-r_cu_2^2}r_c\ve u_{2\perp}
\end{array}\right.
\end{eqnarray}
in which $\rho$ is the density including the protons and electrons, $\ve B$ is the magnetic field with the
subscript `$n$' (`$\perp$') parallel (perpendicular) to the shock normal, $\ve u$ is the solar wind speed
in the DeHoffman-Teller (HT) frame, and $v_A=\sqrt{\frac{B^2}{\mu\rho}}$ is the Alfv{\'e}n speed. The recovered
interval is longer than the shocked interval, and its duration is calculated by using the formula
\begin{eqnarray}
d t_1=\frac{u_{2n}+v_s}{u_{1n}+v_s}r_cd t_2 \label{eq:dt}
\end{eqnarray}
based on the mass conservation. The recovered parameters are plotted
in Figure~\ref{fig:shock_recover}.

The shock parameters, $\hat{\ve n}$, $v_s$ and $r_c$, are obtained by using a nonlinear least-squares fitting
technique~\citep{Vinas_Scudder_1986, Szabo_1994} based on the incomplete Rankine--Hugoniot conditions. % (temperature information is not used).
A total of 10 data points with time resolution of 92 s between 03:00:18 UT and 03:18:11 UT on February 29 are used in the fitting.
The calculated shock normal is ($-0.93$, $-0.01$, $-0.37$) in GSE coordinates, the shock speed is $v_s=585$ km s$^{-1}$ in
the spacecraft frame, and the compression ratio is $1.69$.

The assumptions in recovering the shocked structure are highly ideal. Particularly, the compression ratio in
the sheath region cannot be the same. To check the influence of the compression ratio on the fitting result
of the magnetic cloud at the Earth, we replace the uniform $r_c$ in the sheath region with a varying
$r_c$ linearly decreasing from $1.69$ at the shock surface to $1.0$ in the following $6.5$-hr duration. Using the same
technique described above, we fit the recovered magnetic cloud. The results, indicated as `$\times$' symbols
in Appendix Figure~\ref{fig:fitting_cratio}, are consistent with those (the square symbols) by using
the uniform $r_c$, and do not change the conclusion we will reach below.
This suggests that the ideal assumptions are acceptable for this study.

\begin{table*}[t]
\begin{center}
%\footnotesize
\caption{Magnetic properties of the magnetic cloud at different distances}\label{tb:fitting}
\tabcolsep 10pt %\renewcommand{\arraystretch}{0.42}
\begin{tabular}{ccccccccc}
\hline
	& $r$ & ($\theta$, $\phi$) & $|d|$ & $F_z$ & $\tau$ & ${\tau_{_{AU}}}$ & ${h_{m,_{AU}}}$ & $D_{im}$\\
	& AU & deg & $R_{_{MC}}$ & $\times10^{20}$ Mx & turns/AU & turns/AU & $\times10^{40}$ Mx$^2$ & \% \\
\hline
	Mercury & $0.35$ & ($-65$, $15$) & $0.18_{-0.10}^{+0.25}$ & $11.0_{-5.3}^{+5.8}$ & $-3.8_{-2.5}^{+1.9}$ & $-1.3_{-0.9}^{+0.7}$ & $-160_{-108}^{+79}$ & $15.2_{-7.2}^{+5.8}$ \\
	Venus   & $0.72(0.84)$ & ($-52$, $333$) & $0.58_{-0.08}^{+0.04}$ & $2.1_{-0.9}^{+1.0}$ & $-6.9_{-2.8}^{+3.5}$ & $-2.4_{-1.0}^{+1.2}$ & $-10.6_{-4.4}^{+6.1}$ & $76.8_{-7.8}^{+8.2}$ \\
	Earth   & $1.0$ & ($-21$, $356$) & $0.54_{-0.11}^{+0.10}$ & $1.0_{-0.6}^{+0.6}$ & $-6.4_{-5.4}^{+3.1}$ & $-6.4_{-5.4}^{+3.1}$ & $-5.1_{-3.0}^{+3.6}$ & $25.2_{-17.2}^{+21.8}$ \\
\hline
\end{tabular}
	Column 2: the heliocentric distance of the planets during the period of interest. The value of $0.84$ in the 
	brackets is the position of the nose of the magnetic cloud when the cloud encountered Venus as shown in Fig.\ref{fig:dips}. Column 3: 
	the orientation, i.e., the elevation and azimuthal angles, of the 
	axis of the magnetic cloud in the planet-solar-orbital coordinate system, i.e., MSO, VSO and GSE coordinates for MESSENGER, VEX, Wind, respectively.
	The uncertainty in the orientation is less than $10^{\circ}$. Column 4: The closest approach of the observational path to the axis of the cloud in units of the radius, $R_{_{MC}}$, of the cloud.
	Column 5: the axial magnetic flux. Column 6: the number of turns per AU of the magnetic cloud field lines. Column 7:
	the corresponding $\tau$ when the magnetic cloud arrives at 1 AU, which is given by $\tau_{_{AU}}=\frac{r}{r_{_{AU}}}\tau$ (see Sec.\ref{sec:res}). 
	Column 8: the magnetic helicity per AU at the distance of 1 AU, given by $h_{m,_{AU}}=\tau_{_{AU}}F_z^2$. Column 9: the degree of the imbalance of the azimuthal flux. 
The values for Mercury and Venus are calculated from equation (\ref{eq:imb2}) and that for the Earth from equation (\ref{eq:imb1}).
\end{center}
\end{table*}

\subsection{Results}\label{sec:res}
Three fitting parameters are investigated to study the changes of the magnetic properties of the magnetic cloud, which
are the axial magnetic flux, $F_z$, the number of turns per AU, $\tau$, and the magnetic helicity per AU, $h_m$.
The axial magnetic flux and total magnetic helicity are two invariant parameters for magnetic clouds if no
reconnection is involved with the surrounding magnetic field. This implies that
$\tau$ and $h_m$ both depend on the length of the axis of the magnetic cloud. 
Thus, to make their values obtained at different distances comparable, we normalize them to
the values when the magnetic cloud arrives at the distance of 1 AU. This
normalization can be easily done under the reasonable assumption that the length of the axis of the magnetic cloud
is proportional to the heliocentric distance, $r$. In other words, we can get the normalized values of $\tau$ and $h_m$ by using
$\tau_{_{AU}}=\frac{r}{r_{_{AU}}}\tau$, $h_{m,_{AU}}=\frac{r}{r_{_{AU}}}h_m=\tau_{_{AU}}F_z^2$, respectively. For the magnetic cloud at Mercury,
Venus and the Earth, the value of $r$ is $0.35$, $0.84$ and $1.0$ AU, respectively. Note that the nose of the magnetic cloud
was at $0.84$ AU when the cloud arrived at Venus based on the DIPS model though Venus located at $0.72$ AU (see Fig.\ref{fig:dips}).

Figure~\ref{fig:fitting}b--\ref{fig:fitting}c show the results of $F_z$ and $-h_{m,_{AU}}$, which fall in the typical range estimated in previous statistical 
studies~\citep{Wang_etal_2015}. The averaged values and uncertainties of these parameters, which are calculated based on all the successful test fittings, 
are marked by the dots with error bars (and also listed in Table~\ref{tb:fitting}). It is found that $F_z$ and $-h_{m,_{AU}}$ generally decrease from Mercury 
to the Earth. The averaged value of $F_z$ at the Earth and Venus is about only $9\%$ and $19\%$ of that at Mercury, respectively. Similarly,
the averaged value of $-h_{m,_{AU}}$ at the Earth and Venus is about only $3\%$ and $7\%$ of that at Mercury. 
If the uncertainty in the fitting parameters are considered, the decreases are still notable, which are at least $28\%$ and $54\%$, respectively, in
$F_z$ and $10\%$ and $19\%$, respectively, in $-h_{m,_{AU}}$. 
In contrast, the derived twists at Mercury are obviously weaker
than (or about $0.2$ times of) those at the Earth as shown in Figure~\ref{fig:fitting}d, and the twists at Venus locate between.

\section{Possible interpretations for the model results}\label{sec:interpretations}

\begin{figure*}[t]
\begin{center}
\includegraphics[width=\hsize]{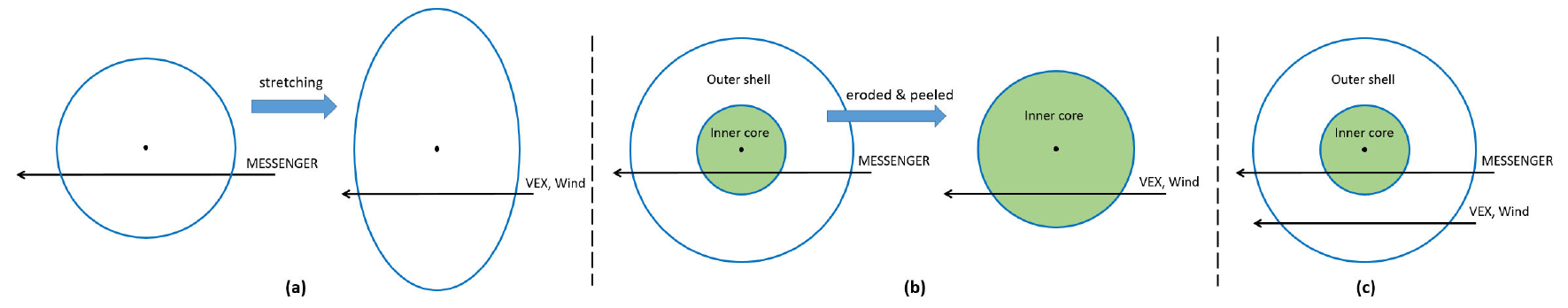}
	\caption{Schematic diagram showing three possible scenarios explaining the variations of the magnetic properties from Mercury to the Earth. In Scenario 1, 
the magnetic cloud was stretched which may cause the axial flux underestimated and twist overestimated. In Scenario 2 and 3,
the magnetic cloud is roughly divided into two parts: an inner core and an outer shell.
Scenario 2 suggests that there was a great erosion during the cloud propagated toward the Earth, and all the spacecraft passed through the inner core. 
Scenario 3 does not include a significant erosion, and only MESSENGER cut through the inner core according to the closest approaches of the observational path to the cloud's axis derived by the fitting method. 
The analysis suggests that Scenario 1 and 2 could explain the observations 
(see Sec.\ref{sec:res} and \ref{sec:erosion} for more details).}\label{fig:scenarios}
\end{center}
\end{figure*}

\subsection{`Pancaking' effect}
There are several possible interpretations for the decreases of the axial magnetic flux and helicity and the increase of the twist
as illustrated in Figure~\ref{fig:scenarios}. First one is due to the `pancaking' effect~\citep[or called stretching effect, e.g.,][]{Crooker_Intriligator_1996, 
Russell_Mulligan_2002, Riley_etal_2003, Riley_Crooker_2004, Manchester_etal_2004a}, 
which makes the cross-section of a MFR deviated away from a circular shape (Fig.\ref{fig:scenarios}a).
Based on the theoretical analysis on the linear force-free field by \citet{Demoulin_Dasso_2009}, it is suggested that 
the axial flux might be underestimated, say by a factor of $a$, if using a cylindrical model to fit a stretched cloud, but 
it will have little effect on the azimuthal flux. As 
a consequence, the ratio of the azimuthal flux per unit length to the axial flux, i.e., a kind of averaged
twist, will be overestimated by a similar factor. 
Thus, the decrease/increase of the axial flux/twist due to 
the `pancaking' effect is not real but from the model bias. 

It should be noted that the twist in our model is not estimated based on the
ratio of the azimuthal flux to the axial flux, but independently obtained by fitting to the measurements of $\frac{B_\varphi}{xB_z}$, in which $B_\varphi$ and $B_z$ are two
components of the magnetic field in the magnetic cloud frame $(r, \varphi, z)$ with $z$ along the axis of the magnetic cloud and $x$ is the distance from the cloud axis normalized by its radius $R_{_{MC}}$ 
(see the description in Sec.2.2 of \citealt{Wang_etal_2016}). We can imagine that the `pancaking' effect can make $R_{_{MC}}$ underestimated by a factor 
of the order of $\sqrt{a}$ but have little to 
do with $\frac{B_\varphi}{B_z}$. Thus, the overestimation factor of the twist value by this method should be smaller than that by using the ratio of the two fluxes.

If the underestimation factor, $a$, in the axial flux was $11$ at the Earth, this
effect could well explain the decrease of the axial flux that is about 91\% from Mercury to the Earth.
However, when reaching the underestimation factor of $11$, the magnetic cloud should have been highly stretched with the aspect
ratio of its cross-section of more than $10$ according to Fig.8 in the paper by \citet{Demoulin_Dasso_2009}.
Some MHD numerical simulations showed that the aspect ratio is about $3$ or less near $1$ AU [see, e.g., Fig.3 in \citealt{Riley_etal_2003} 
and Fig.5 in \citealt{Manchester_etal_2004a}].
Other simulations suggested that the `pancaking' effect is not so significant even if a magnetic cloud is compressed 
by a following fast shock and/or ejecta [see Fig.3 in \citealt{Xiong_etal_2006a} and Fig.1 in \citealt{Xiong_etal_2007}].
Assuming that the aspect ratio of the stretched cross-section of the cloud is $3$, which is large enough according to those simulations, 
we may read from Fig.8 of \citet{Demoulin_Dasso_2009} that
the underestimation factor of the axial flux is about $3.2$, leading to its apparently decrease by about $69\%$ from Mercury to the Earth,
marginally explaining the derived decrease of the axial flux if the uncertainties in the derived fluxes are considered. Similarly, the increase
of the twist from Mercury to the Earth may also be marginally explained by the `pancaking' effect with the uncertainties considered.

\begin{figure*}[t]
\begin{center}
\includegraphics[width=\hsize]{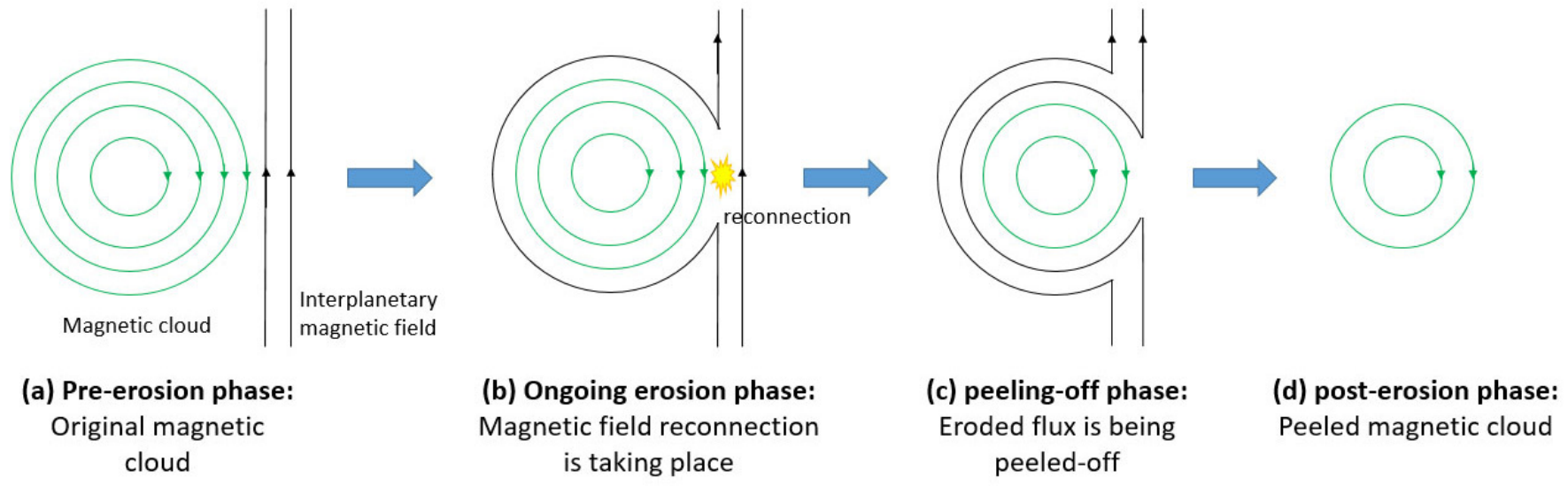}
	\caption{Schematic diagram showing four phases of a complete erosion process of a magnetic cloud: (a) pre-erosion phase, (b) ongoing erosion phase, (c) peeling-off phase and (d) post-erosion phase.}\label{fig:erosion}
\end{center}
\end{figure*}

\subsection{`Erosion' effect}
It was suggested that magnetic clouds may experience erosion process~\citep[e.g.,][]{Dasso_etal_2006, Ruffenach_etal_2012, 
Ruffenach_etal_2015, Manchester_etal_2014} 
through the magnetic reconnection with ambient solar wind~\citep{Gosling_2012} when they propagate away from the Sun.
A previous statistical study of $50$ magnetic clouds~\citep{Ruffenach_etal_2015} showed that up to $90\%$ of magnetic flux, with an 
average of $40\%$, can be eroded based on the imbalance of azimuthal magnetic flux in these clouds. 
A complete erosion process roughly consists of four phases as illustrated in Figure~\ref{fig:erosion}: a pre-erosion phase, during which the magnetic field lines of a magnetic cloud are not reconnected with the magnetic field lines
in the ambient solar wind yet; an ongoing erosion phase, when the reconnection is taking place; a peeling-off phase, when the reconnected field lines are being peeled off from the magnetic cloud; a post-erosion phase, the eroded magnetic field flux has been completely peeled-off from the magnetic cloud. The second and third phases may happen simultaneously.
Figure~\ref{fig:scenarios}b shows an example of erosion by dividing the magnetic cloud into two parts: an inner core and an outer shell.
The outer shell is gradually eroded during the propagation. The observational signature of the erosion of this event will be given later in Sec.\ref{sec:erosion}.
Here we will see if this scenario can explain the decrease of the axial flux and the increase of the twist and 
if it is consistent with the observed profile of magnetic field from the spacecraft.

To better understand this scenario, the closest approach, $d$, of the 
observational path to the axis of the magnetic cloud derived from the fitting method is listed in Table~\ref{tb:fitting} for reference. 
It is suggested that the MESSENGER spacecraft at Mercury was relatively much closer to the axis of the 
cloud than VEX at Venus and Wind at the Earth. 
Thus, all the spacecraft passed through the inner core of the magnetic cloud.
%Both of the two scenarios may result in the observed decreases of the axial magnetic flux and helicity and the increase of the twist, 
%but the profile of the magnetic field from the axis of the magnetic cloud to the boundary will be quite different between the two scenarios. This could be seen from the following quick estimation.
Based on Figure~\ref{fig:scenarios}b, we may assume that the boundary of the inner core initially locates between $0.2R_{_{MC}}$ and $0.5R_{_{MC}}$, 
say at about $0.4 R_{_{MC}}$. 
Moreover, we assume that the magnetic fields in the inner core and the outer shell are roughly constant, setting to be
$\ve B_{core}$ and $\ve B_{shell}$, respectively. 
Then, the axial and poloidal magnetic fluxes and the twist derived from our uniform-twist flux rope model can 
be approximated as 
\begin{eqnarray}
&&\left\{\begin{array}{l}
F_{z,_M}=2\pi\left[B_{core,z}R_{core}^2+B_{shell,z}(R_{_{MC}}^2-R_{core}^2)\right]\\
F_{\varphi,_M}=\left[B_{core,\varphi}R_{core}+B_{shell,\varphi}(R_{_{MC}}-R_{core})\right]L\\
\tau_{_M}=\frac{F_{\varphi,_M}}{F_{z,_M}L}=\frac{B_{core,\varphi}R_{core}+B_{shell,\varphi}(R_{_{MC}}-R_{core})}{F_{z,_M}}
\end{array}\right.\label{eq:par1_m}
\end{eqnarray}
if the spacecraft crossed the cloud with the closest approach like MESSENGER, and approximated as
\begin{eqnarray}
&&\left\{\begin{array}{l}
F_{z,_E}=2\pi(B_{core,z}R_{core}^2)\\
F_{\varphi,_E}=B_{core,\varphi}R_{core}L\\
\tau_{_E}=\frac{F_{\varphi,_E}}{F_{z,_E}L}=\frac{B_{core,\varphi}R_{core}}{F_{z,_E}}
\end{array}\right.\label{eq:par1_e}
\end{eqnarray}
if the spacecraft crossed the cloud like Wind. Here $L$ is the length of the axis of the cloud. 
%The formulae of the parameters at Venus are similar to those at the Earth, but we do not need them, 
%because the comparison of the parameters between the cloud at Mercury and the Earth is enough to reveal the 
%difference in the magnetic field profile between the two scenarios.
Based on our model results (see Table~\ref{tb:fitting}), we roughly have $\frac{F_{z,_M}}{F_{z,_E}}\approx10$ and 
$\frac{\tau_{_M}}{\tau_{_E}}\approx0.2$. From equations (\ref{eq:par1_m}) and (\ref{eq:par1_e}), we can deduce that
$B_{core}\approx\sqrt{(1.5B_{shell,\varphi})^2+(0.58B_{shell,z})^2}$, or $0.58B_{shell}<B_{core}<1.5B_{shell}$.
It suggests that the magnetic field is flattened from the inner core to outer shell, consistent with 
the magnetic field profile measured by MESSENGER as shown
in the first panel of Figure~\ref{fig:mc_mer_ven}a. Thus, this scenario can also explain the derived variations in the axial flux,
helicity and twist, and differently from the `pancaking' effect, these variations are real.

\subsection{Double-layer structure without erosion}
Another possible scenario is as shown in Figure~\ref{fig:scenarios}c, in which the magnetic cloud is also considered as a combination of 
an inner core and an outer shell as the previous scenario. But in this case, there was no significant erosion happening to the cloud, and 
VEX and Wind only cut through the outer shell of the magnetic cloud in contrast to MESSENGER which crossed through its inner core.
This scenario might also explain the decrease of the axial flux and increase of the twist if the inner core carries a stronger magnetic field
and a weaker twist than the outer shell. However, the similar analysis of the values of $B_{core}$ and $B_{shell}$ presented below disapproves the possibility.

In this scenario, equations (\ref{eq:par1_e}) should be revised as
\begin{eqnarray}
&&\left\{\begin{array}{l}
F_{z,_E}=2\pi(B_{shell,z}R_{_{MC}}^2)\\
F_{\varphi,_E}=B_{shell,\varphi}R_{_{MC}}L\\
\tau_{_E}=\frac{F_{\varphi,_E}}{F_{z,_E}L}=\frac{B_{shell,\varphi}R_{_{MC}}}{F_{z,_E}}
\end{array}\right.\label{eq:par2_e}
\end{eqnarray}
and the relation between $B_{core}$ and $B_{shell}$ becomes
$B_{core}\approx\sqrt{(3.5B_{shell,\varphi})^2+(57B_{shell,z})^2}$, or $3.5B_{shell}<B_{core}<57B_{shell}$, 
suggesting a much stronger magnetic field in the inner core than in the outer shell. 
It does not match the magnetic field profile measured by MESSENGER or Wind.

\begin{figure*}[t]
\begin{center}
\includegraphics[width=\hsize]{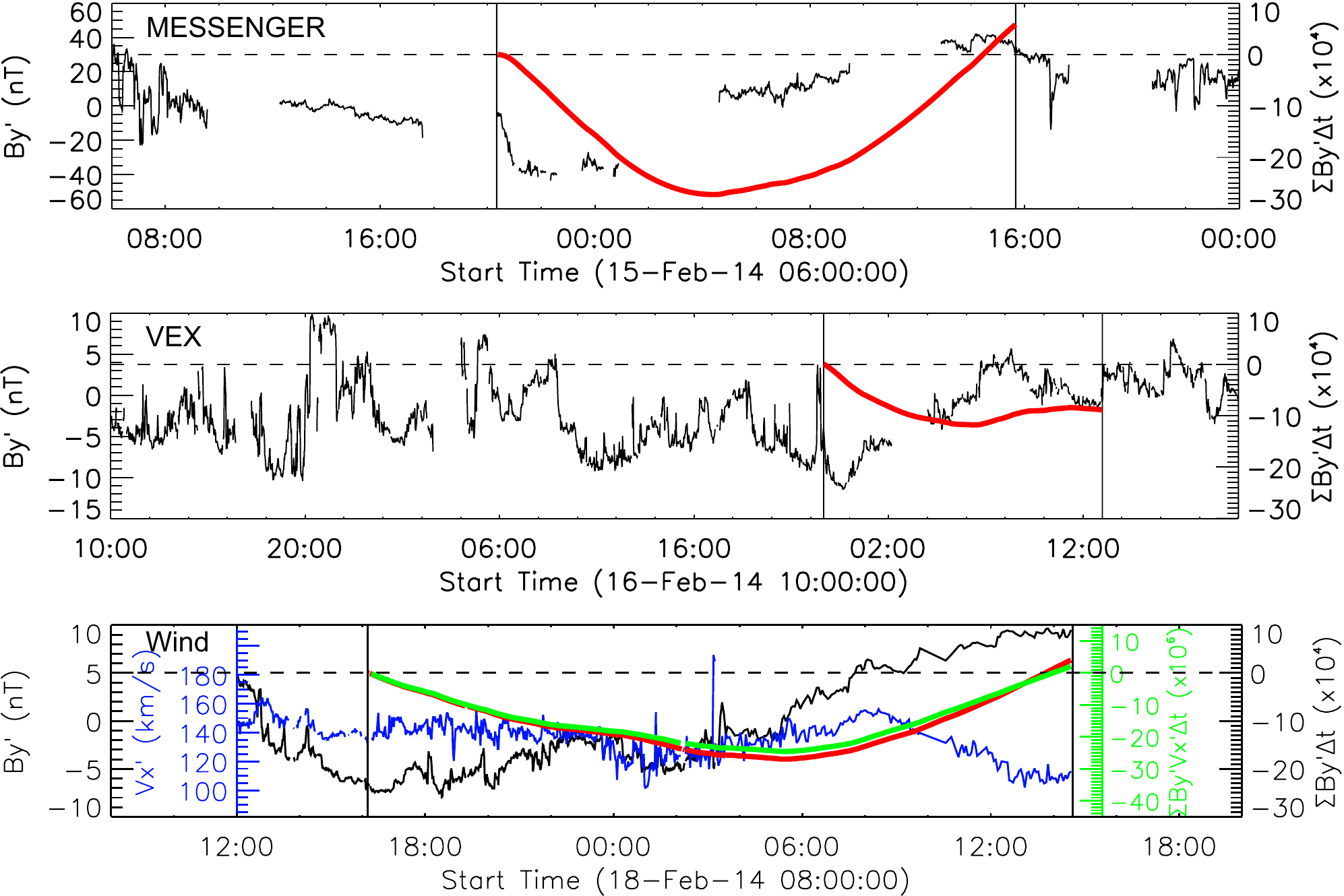}
	\caption{Imbalance of the azimuthal magnetic flux. The black curves in all the panels show the profiles of the $\ve y'$
	component of the magnetic field in the magnetic cloud frame ($\ve x'$, $\ve y'$, $\ve z'$) with $\ve z'$ along the axis of the magnetic 
	cloud and $\ve y'$ perpendicular to the plane defined by $\ve z'$ and the observational path of the spacecraft. The blue curve in 
	the last panel shows the profile of the $\ve x'$ component of the solar wind velocity, and is scaled by the second vertical axis on the left. The vertical lines mark the boundary 
	of the magnetic cloud. The thick red lines give the profile of the cumulative value of $B_{y'}$ with the time by using 
	equation (\ref{eq:imb2}), and the thick green line the profile of the cumulative value of $B_{y'}v_{x'}$ by using equation (\ref{eq:imb1}). 
All these thick lines have been corrected to the values when the cloud arrives at 1 AU by multiplying the distance ratio $\frac{r}{r_{_{AU}}}$ as what we did to $h_m$ and $\tau$, and 
the scales are given by the vertical axes on the right in the units of (nT s) for the red lines or (nT km) for the green line. 
	The deviation of the right ends of these thick lines away from 
	zero indicates a possible imbalance.}\label{fig:imbalance}
\end{center}
\end{figure*}

\subsection{Signatures of the erosion process possibly experienced by the magnetic cloud}\label{sec:erosion}

Both the `pancaking' and `erosion' effects may explain the variations of the derived magnetic properties.
However, it is difficult to assess how significant the `pancaking' effect was based on the one-dimensional 
data. Here, we focus on the erosion effect to look for observational signatures. A frequently used signature is the 
imbalance of azimuthal magnetic flux of magnetic clouds.
The azimuthal magnetic flux is calculated in the magnetic cloud frame ($\ve x'$, 
$\ve y'$, $\ve z'$) with $\ve z'$-axis along the orientation of the axis of the cloud and $\ve y'$-axis perpendicular to both $\ve z'$-axis and the observational
path of the spacecraft. The measured magnetic field and solar wind velocity are then projected onto the ($\ve x'$, $\ve y'$) plane.
For a complete MFR, the azimuthal magnetic flux cumulated from one boundary of the 
MFR to the other along the observational path should be zero. A deviation from zero is the imbalanced flux, $F_{im}$, estimated as
\begin{eqnarray}
	\frac{F_{im}}{L}=\int_{in}^{out} B_{y'}v_{x'}dt \label{eq:imb1}
\end{eqnarray}
in which $L$ is the length of the MFR, $B_{y'}$ and $v_{x'}$ is the measured magnetic field and solar wind speed along the 
$\ve y'$ and $\ve x'$ directions, respectively, in the magnetic cloud frame, and `{\it in}' and `{\it out}' indicate 
the integral through the front boundary of the cloud to the rear boundary. 
The imbalance of azimuthal flux provides evidence of eroded but not yet peeled-off flux (i.e., in the second and third phases of the erosion process, see Fig.\ref{fig:erosion}), but may miss the completed erosion in which 
the flux has been completely peeled off. 

Based on the orientation obtained from the fittings (see Table~\ref{tb:fitting}), 
we convert the magnetic field components into the magnetic cloud frame. The profiles of $B_{y'}$ recorded by MESSENGER,
VEX and Wind are shown in Figure~\ref{fig:imbalance}. Wind spacecraft has valid measurements of solar wind velocity, 
and therefore the profiles of $v_{x'}$ is plotted in the last panel. Since there is no valid measurements of solar 
wind velocity in the MESSENGER and VEX data, we simply assume that the magnetic
cloud was uniformly propagating through Mercury and Venus. The imbalance of the flux can be evaluated by a revised formula
\begin{eqnarray}
	\frac{F_{im}}{v_{x'}L}=\int_{in}^{out} B_{y'}dt \label{eq:imb2}
\end{eqnarray} 
The data gaps in the measurements are filled by the linear interpolation.
The red curves in Figure~\ref{fig:imbalance} are calculated according to equation (\ref{eq:imb2}). 
For the magnetic cloud at the Earth, the data of the recovered uncompressed structure are used. The green curve in the last panel
is calculated by equation (\ref{eq:imb1}). The red curves in the first two panels are all corrected to the values when the cloud arrives at 1 AU by applying a factor of $\frac{r}{r_{_{AU}}}$.
In the last panel, the two curves have similar shapes, 
suggesting that the red curves by equation (\ref{eq:imb2}) in the other two panels should be reliable.

\begin{figure}[hb]
\begin{center}
\includegraphics[width=\hsize]{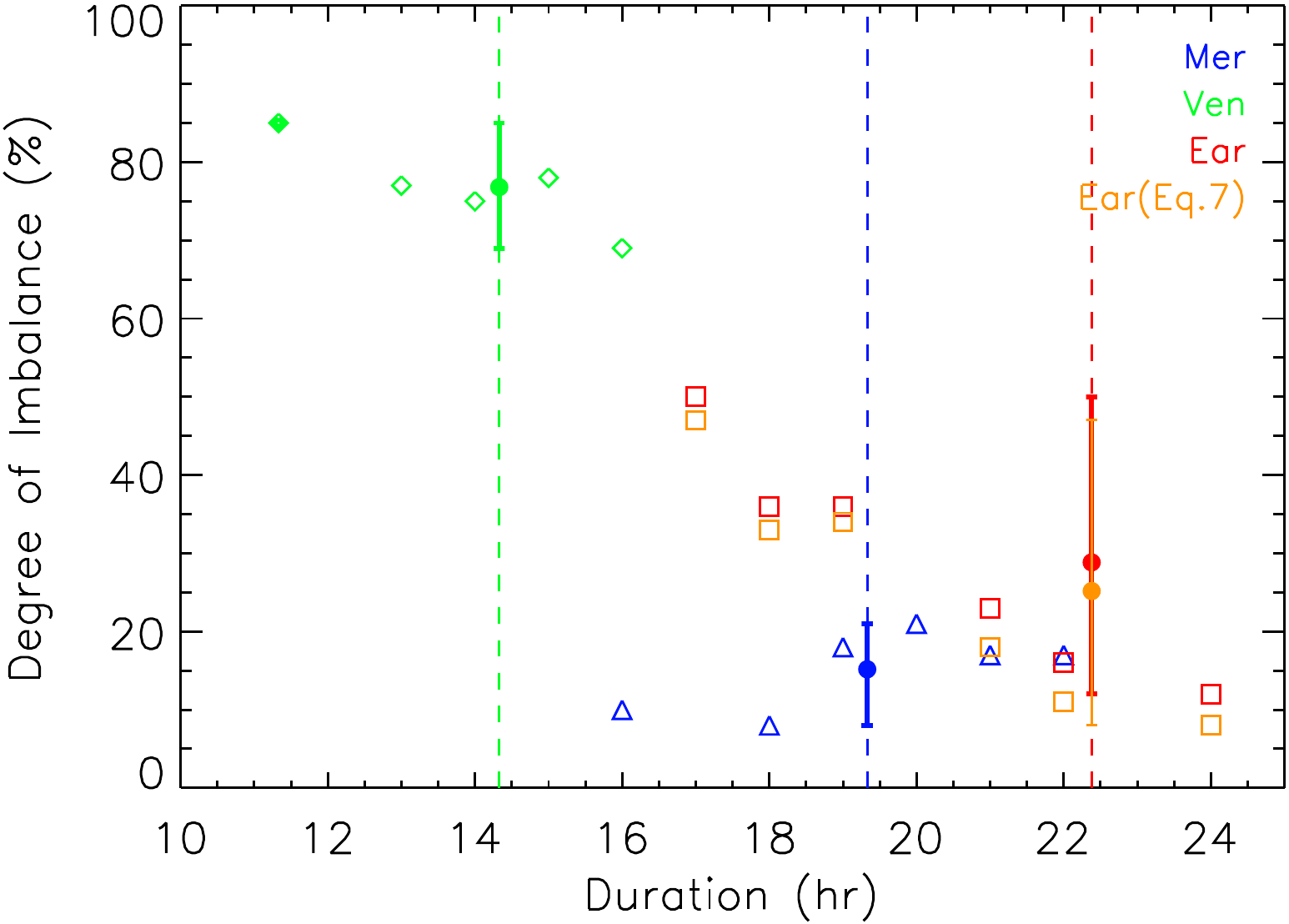}
\caption{Degree of the imbalance of the azimuthal flux for the cloud with different boundaries. The symbols follow the same meaning as those
in Fig.\ref{fig:fitting}. For the cloud at Venus, we add an additional test case by moving front boundary of the
cloud inward by $3$ hours as shown by the filled green diamond. The imbalances at Earth calculated based on equation (\ref{eq:imb2}) and equation (\ref{eq:imb1}) are 
displayed by the symbols in red and orange, respectively.}\label{fig:imbalance-duration}
\end{center}
\end{figure}

It can be seen that an imbalance in the azimuthal magnetic flux can be found at all the three distances and their significances are different.  
The degree of the imbalance, defined as the ratio of the imbalanced flux to the total flux, is less than 18\% at Mercury and the Earth
reading from the imbalance curves in the top and bottom panels, and about 75\% at Venus. To test the effect of choosing the boundaries on the imbalance, we adjust the 
boundaries of the cloud by using the same aforementioned method and derive the 
degree of the imbalance as shown in Figure~\ref{fig:imbalance-duration} (also listed in the last column of Table~\ref{tb:fitting}). 
It is found that in our test cases, the degree of the imbalance is small, about $15\%$ at Mercury and then increases to about $77\%$ at Venus and $25\%$ at the Earth. 
The uncertainty in the imbalance degree at the Earth is quite large, which suggests that the degree might reach up to about $50\%$.
Thus, the erosion effect did exist in this event and probably contributed to the variations of the derived axial 
flux and twist with the changing heliocentric distance. The difference of the imbalance degree among the three locations might be due to (1) the 
model errors, (2) that the erosion and peeling-off processes continued to progress between Mercury and the Earth, and/or (3) that some eroded flux has been 
completely peeled off at some locations and therefore not taken into account by this method. 
For an ongoing erosion process, magnetic reconnection should happen somewhere at the boundary of the magnetic cloud. As to this event, we 
do not find any significant signatures of reconnection, implying that the spacecraft probably did not cross the reconnection region.

\section{Summary and discussion}

\begin{figure*}[t]
\begin{center}
\includegraphics[width=\hsize]{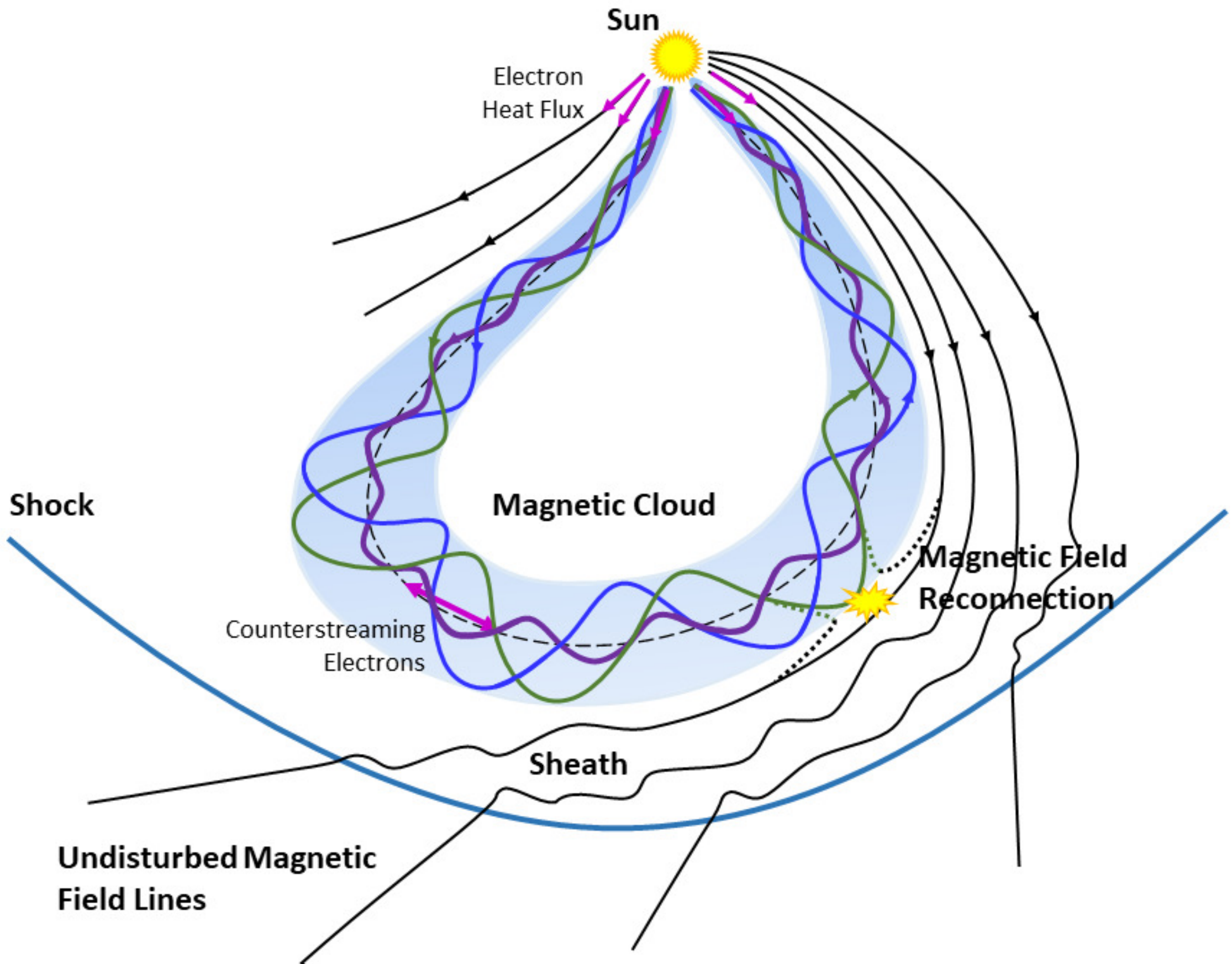}
	\caption{A cartoon showing a typical magnetic cloud in the heliosphere, redesigned based on the picture by 
	\citet{Zurbuchen_Richardson_2006}. The inner purple field line is more twisted than other two outer
	lines in the magnetic cloud. The reconnection site implies the erosion process.}\label{fig:picture}
\end{center}
\end{figure*}

%With the aid of the uniform-twist force-free flux rope model, a magnetic cloud observed by multiple spacecraft well-aligned along the radial
%direction is studied. 
In this study, we investigate a magnetic cloud propagating through Mercury, Venus, Earth and Mars. The magnetic cloud was overtaken by 
a following fast ejecta and the ejecta-driven shock near the Earth and caused a Forbush decrease at Mars. A method to recover a shock-compressed structure
is developed and applied to the magnetic cloud observed by the Wind spacecraft at $1$ AU. With the aid of the uniform-twist force-free
flux rope model, the axial magnetic flux, helicity and twist per unit length of the magnetic cloud were derived at three heliocentric distances:
Mercury, Venus and the Earth. It is found that the axial flux and helicity decreased from Mercury to the Earth but the twist increased. 

Two effects may be responsible for these variations with the heliocentric distance, the `pancaking' effect and the `erosion' effect. Our analysis combined 
with previous simulations and theoretical analysis~\citep[e.g.,][]{Riley_etal_2003, Manchester_etal_2004a, Xiong_etal_2006a, Xiong_etal_2007, Demoulin_Dasso_2009} 
suggests that the `pancaking' effect may marginally explain the phenomena if the initially cylindrical magnetic cloud was distorted and stretched 
to a nearly pancake shape with the aspect ratio of its cross-section being as large as $3$. However, based on the present one-dimensional data, it is difficult to
estimate how significant the `pancaking' effect is for this magnetic cloud. In this scenario, the variations in the axial flux, helicity
and twist do not mean real changes of these properties of the magnetic cloud, but come from the model bias when the shape of the cross-section deviates from the cylindrical model.

The erosion effect is evident by the imbalance of the azimuthal magnetic flux at all the three locations: Mercury, Venus and the Earth. The degree of
the imbalance at Mercury, Venus and the Earth is about $15\%$, $77\%$ and $25\%$, respectively. Although the imbalance degree at Mercury and the Earth is 
less significant than that at Venus, it does suggest that an erosion process was taking place.
%, and it is possible that there were some fluxes completely peeled-off
%during the propagation of the magnetic cloud from Mercury to 1 AU that probably was not taken into account in our analysis. 
This erosion effect may stand alone to 
explain the variations of the axial flux, helicity and twist. In this scenario, these variations are real and imply that the magnetic cloud consists of
a high-twist core and a weak-twist outer shell. However, again, we cannot exclude the possibility of the `pancaking' effect, which more or less happens to 
magnetic clouds in interplanetary space. Thus, as a conclusion, it is likely that both effects jointly caused such variations with the heliocentric distance.

Since erosion effect exists and the twist increase is real in case of this effect, we would like to discuss its implications on the formation of MFRs.
First, the erosion process caused the inner core of the magnetic cloud exposed in the solar wind at far distance. 
As mentioned before, it leads to the possibility that
the twist in the cross-section of the initial magnetic cloud was non-uniform, but roughly stage-like distributed with a high-twist core
inside. The global picture of an interplanetary magnetic cloud~\citep{Zurbuchen_Richardson_2006} may be further modified as 
Figure~\ref{fig:picture}, in which the elements of the stage-like twist distribution and an erosion process are incorporated. 

Second, back to the debates 
mentioned at the beginning, if the `pancaking' effect was insignificant as we argued here, the event presented in this paper supports the scenario 
that a seed MFR probably exists prior to the CME eruption, and the magnetic 
field lines added through the magnetic reconnection during the eruption constitute the outer flux with a twist less
than the inner seed MFR. %Though we lack the information of the source region of the magnetic cloud, 
Regretfully, the magnetic cloud was a slow and therefore weak one. Its corresponding CME is difficult 
to be distinguished from other preceding and following CMEs during the period, and its source location is ambiguous (see Appendix~\ref{supp:mc_source}). 
Thus, we cannot find more supporting material from its source region for this event. But 
previous studies have showed the possibility of preexisting seed 
MFRs~\citep[e.g.,][]{Chintzoglou_etal_2015, LiuR_etal_2016}, which are thought to be the necessary condition of a successful eruption~\citep{LiuL_etal_2016}, 
based on solar multiple-wavelength observations. The recent theoretical work by \citet{Priest_Longcope_2017} also suggested that no high-twist 
core can form without a preexisting MFR.

Besides, the picture of magnetic field lines possessing a strong twist in the core of a MFR but a weak twist in the outer shell is consistent with the relation of 
$\Phi_c=2\frac{l}{R}$~\citep{Dungey_Loughhead_1954, Wang_etal_2016}, implying that the outer magnetic field lines twist weaker and weaker when a MFR grows up in terms of
the kink instability. 
Such a stage-like distribution of twist in magnetic clouds was roughly revealed by 
the Grad-Shafranov reconstruction of magnetic clouds~\citep{Hu_etal_2015}, and was also showed in the most recent observational work on a solar MFR~\citep{WangW_etal_2017}.
Although the study presented here does not yet reach a definite conclusion about the twist distribution inside the MFR due to the presence of the `pancaking' effect,
we do bring additional insights to the formation and internal structure of MFRs from a unique angle of view.
The upcoming space missions `Parker Solar Probe' and `Solar Orbiter' will provide more opportunities for anatomizing an interplanetary magnetic cloud at different 
distances by multiple radially-aligned spacecraft, and the analysis methodology established in this study will show its merits.

\begin{acknowledgments}
We acknowledge the use of the data from the magnetometers onboard the MESSENGER, VEX and Wind spacecraft, 
	the Solar Wind Experiment (SWE) onboard Wind spacecraft, the RAD onboard the MSL, and EUV imagers and 
	coronagraphs onboard the Solar Dynamics Observatory (SDO), Solar and Heliospheric Observatory (SOHO) and the twin
	Solar Terrestrial Relations Observatories (STEREO). We do appreciate the constructive comments from anonymous referees.
        which make the paper much better. Y.W. are grateful to the valuable discussion
	with Hui Li from Los Alamos National Laboratory about the astrophysical jets and high twist phenomena. 
	J.G. acknowledges stimulating discussions with the ISSI team ``Radiation Interactions at Planetary Bodies'' 
	and thanks ISSI for its hospitality. Y.W. acknowledges the support from NSFC grants 41774178 and 41574165, 
	R.L. the support from NSFC grants 41474151 and 41774150 and the Thousand Young Talents Program of China, 
	J.L. the support from the Science and Technology Facility Council (STFC) of UK, and 
	Q.H. partial support from NASA grant NNX14AF41G and NRL contract N00173-14-1-G006.
	This work is also supported by NSFC grants 41761134088 and 41421063 and the fundamental research funds for the central universities.
\end{acknowledgments}

%\clearpage
\appendix

\section{The corresponding CMEs of ejecta `E1', `E3' and `E4' at Earth and their counterparts at Mercury}\label{supp:corres_cmes}

\begin{figure*}[hb]
\begin{center}
\includegraphics[width=\hsize]{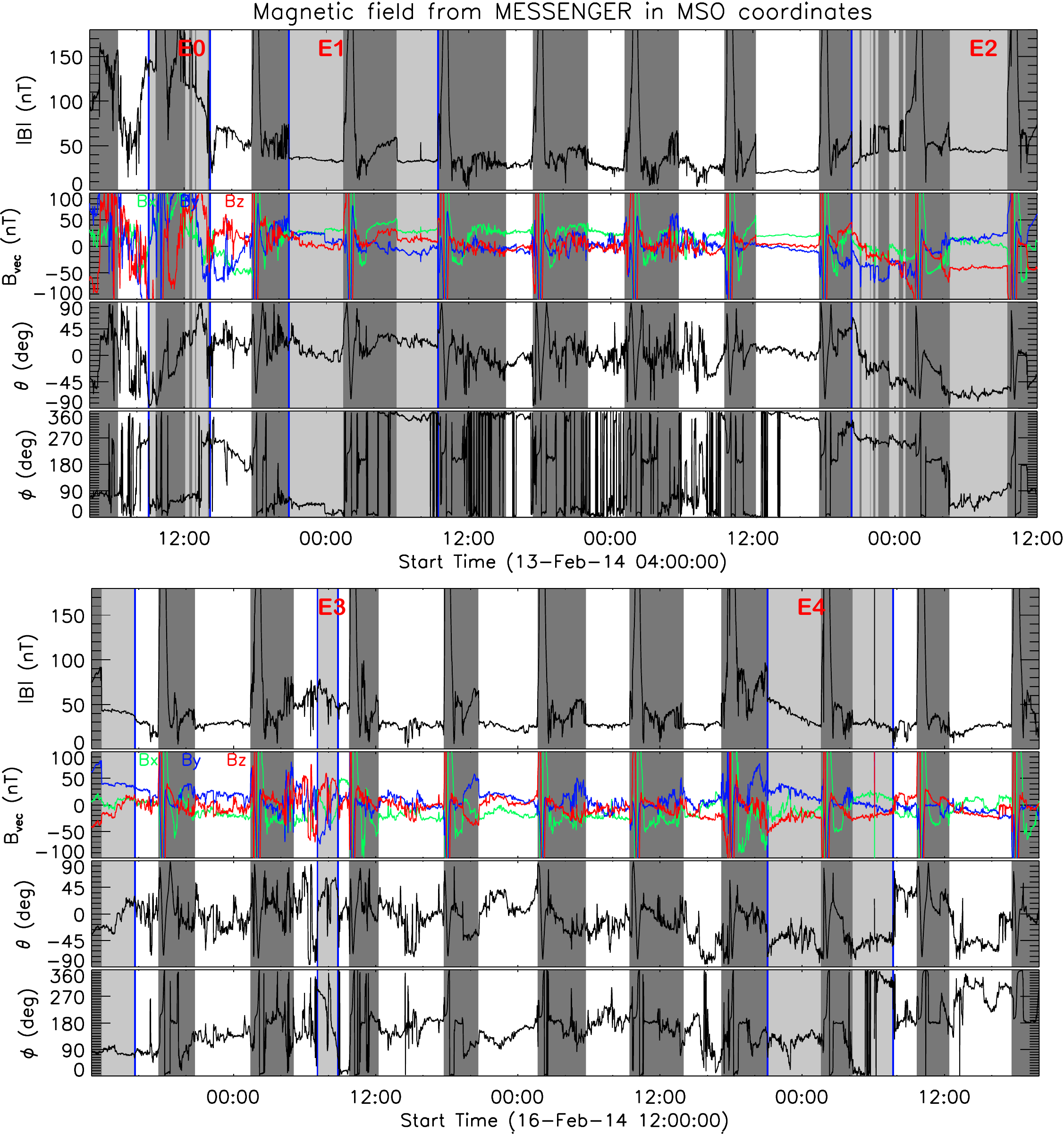}
	\caption{Magnetic fields measured by MESSENGER from February 13 04:00 UT to 19 20:00 UT. The arrangement is 
	the same as that in Fig.\ref{fig:mc_mer_ven}a.}\label{fig:cmes_mer}
\end{center}
\end{figure*}

The ejecta `E1', `E3' and `E4' observed at Earth can be found their counterparts at Mercury.
Figure~\ref{fig:cmes_mer} shows the magnetic field during February 13 -- 19. Except for the magnetic cloud already
studied, we can identify other four ejecta, as indicated by light-shadowed regions bounded by vertical blue lines.
In all of these regions, the magnetic fields were less fluctuated than ambient magnetic fields and the rotations of field
vectors were clear. According the time sequence, we label them as `E0' through `E4', including the magnetic cloud of interest.
Ejecta `E3'
is much smaller than `E1', `E2' and `E4', but its magnetic field is stronger than theirs. Thus, `E3' may continuously
expand on its way out to reach a reasonable size at Earth. The arrival
times of the front boundaries of these ejecta are listed in Table~\ref{tb:associations}. To verify the associations of
these ejecta with those at Earth, we calculate their transit speeds, $v_{me}$, from Mercury to Earth, which
are also listed in Table~\ref{tb:associations}. It is found that the transit speeds of `E3' and `E4' are well consistent
with the in-situ speeds of the two ejecta observed at Earth. The association of `E1' is also acceptable though its
transit speed is about $60$ km s$^{-1}$ less than its in-situ speed. This difference in speed is not too large,
considering a possible acceleration due to the interactions of the ejecta with ambient solar wind and also with the following faster ejecta.

Ejecta `E0' is right ahead of `E1' in the MESSENGER data, which carried a strong magnetic field.
This ejecta cannot be associated to `E1' at Earth, because the transit speed would be even lower than expected. We
check again the data from the Wind spacecraft, and find that there was indeed an evident magnetic cloud with an in-situ
speed of about $380$ km s$^{-1}$ arriving at Earth at 04:05 UT on February 16 (figure is not shown here, but can be
found at our website \url{http://space.ustc.edu.cn/dreams/wind_icmes/}), quite consistent with the transit speed of
about $400$ km s$^{-1}$. In all of these five ejecta observed at Mercury, `E2' demonstrates more typical features of
a magnetic cloud than others. That is why we choose `E2' as the target in this study. It should be noted that
only two of the five ejecta, `E0' and `E3', are listed in the catalog compiled by \citet{Winslow_etal_2015}
based on the MESSENGER data. We confirm the other three not only based on the features in the magnetic field
observed by MESSENGER but also according to the consistent associations between the ejecta at Earth and Mercury.
Besides, due to the $20^\circ$
separation of Venus away from the Sun-Mercury-Earth line, we do not try to make one-to-one associations for these ejecta,
which have made the inner heliosphere much disturbed and complicated.

\begin{figure*}[ht]
\begin{center}
\includegraphics[width=\hsize]{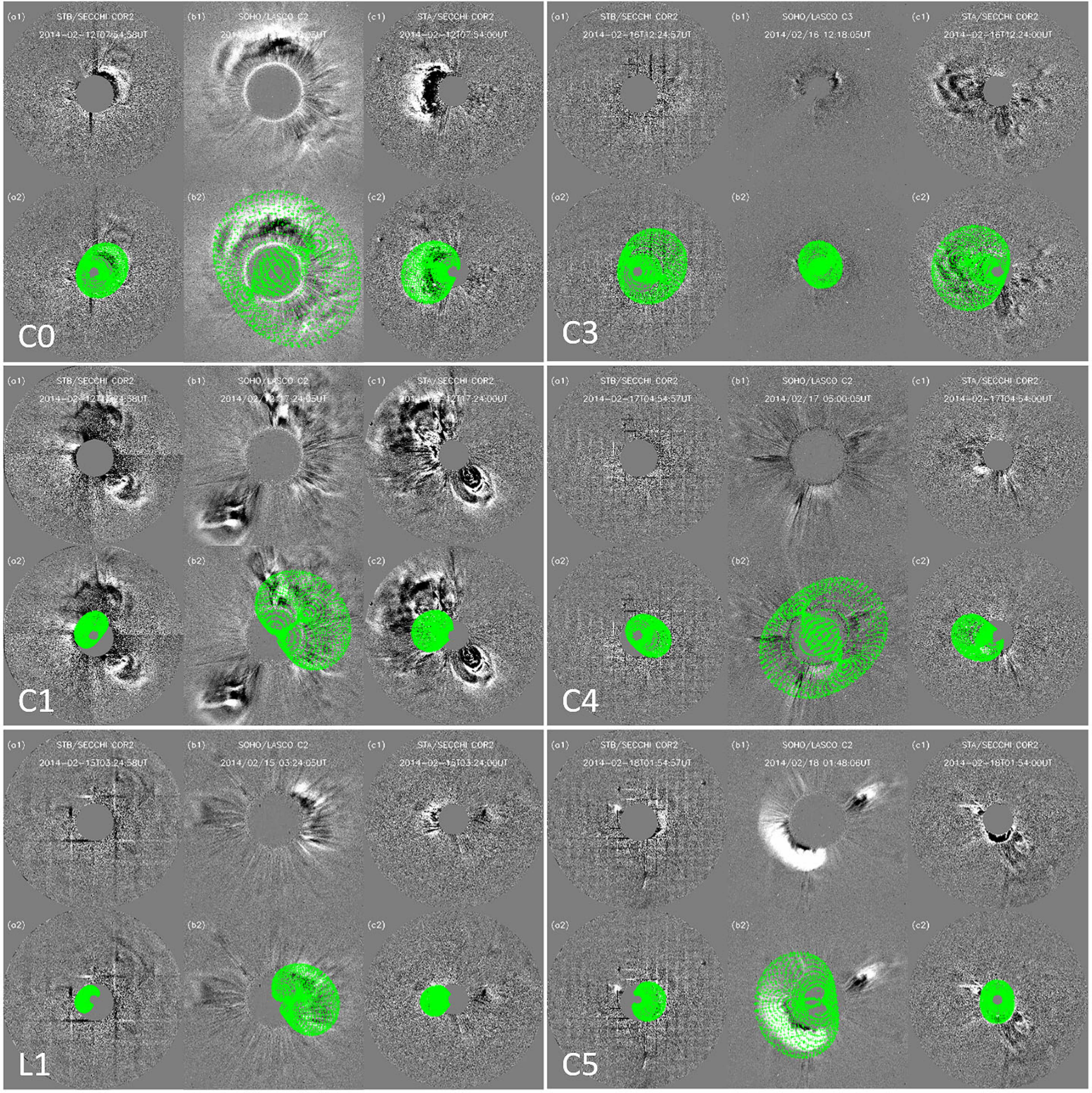}
	\caption{Coronagraph images of CMEs `C0' through `C5' and `L1' (see Table~\ref{tb:associations}), and the same images with GCS fitting meshes (green lines) 
	superimposed on. For each panel, from the left to the right column, it shows the image taken by STEREO-B, SOHO and STEREO-A, 
	respectively.}\label{fig:gcs}
\end{center}
\end{figure*}

The associations of these ejecta with the CMEs observed in coronagraphs are further identified.
The Sun was very productive in February of 2014.
According to the CME catalog~\citep{Yashiro_etal_2004}
compiled based on the observations of the Large Angle and Spectroscopic COronagraph (LASCO, \citealt{Brueckner_etal_1995}) onboard the
Solar and Heliospheric Observatory (SOHO) and our own manually check
with the coronagraph images from COR2s of the SECCHI packages~\citep{Howard_etal_2008} onboard the STEREO-A and B and 
the images from SOHO/LASCO, there were 16 CMEs with apparent angular width larger than $90^\circ$ as listed in Table~\ref{tb:associations}.
Not all of them directed to Earth. By combining the images from SOHO/LASCO and STEREO-A and B/COR2s, we can roughly
determine the propagation directions of these CMEs. The position of the three spacecraft
can be found in Figure~\ref{fig:dips}. It is found that only CMEs labeled as `C0' through `C5' and `L1'
are candidates. `C2' is not listed in the LASCO CME catalog, and we think it is the most probable candidate of the magnetic cloud
of interest in this study, which will be discussed in the next section. Here we
focus on the rest. To get more accurate kinematic parameters of these CMEs in three-dimensional space, we apply a
forward modeling to the coronagraph images with the aid of Gradual Cylindrical Shell (GCS) model~\citep{Thernisien_2011}.
The modeled parameters, which correspond to the CME's leading edge at 20 solar radii, are listed in Table~\ref{tb:associations}.
The meshes fitting to the outlines of these CMEs are shown in Figure~\ref{fig:gcs}.

CMEs `C0', `C3'--`C5' and `L1' can be well recognized in the coronagraphs onboard the SOHO and STEREO-A and B, and therefore
can be well fitted. In these CMEs, `L1' is a limb event viewed from Earth. Its speed was close to the ambient solar wind,
and therefore no significant deflection is expected~\citep{Wang_etal_2016a}. Thus, this CME should not be able to encounter
Mercury and Earth. The other CMEs `C0', `C3'--`C5' are thought to be responsible for ejecta `E0', `E3' and `E4'. CME
`C5' was particularly fast. Though it propagated west to the Sun-Earth line in the corona, it may be deflected toward the Sun-Earth
line in interplanetary space according to our DIPS model~\citep{Wang_etal_2016a}. Thus, it was able to catch up with the preceding one
`C4' and formed a complex ejecta at Earth.
It is noteworthy that all of these CMEs had a faster speed than the transit speed from Mercury to Earth. This phenomenon is
reasonable as CMEs will be quickly assimilated to the ambient solar wind in terms of speed~\citep{Gopalswamy_etal_2000}.
The GCS fitting of CME `C1', however, is not confident, because the CME followed another one, which made it very blurry, especially
in the field of view of the STEREO-B/COR2. Based on the current GCS fitting, the CME initially propagated along the direction $30^\circ$ away
from Earth and might be deflected toward the Sun-Earth line in interplanetary space to encounter Mercury and Earth with its flank.

\begin{table*}[ht]
\begin{center}
\small
\caption{Associations of CMEs in the corona and ejecta at Mercury to the ejecta observed at Earth}\label{tb:associations}
\tabcolsep 1pt
\renewcommand{\arraystretch}{1}
\begin{tabular}{l|cc|cc|cccc|ccc|c}
\hline
          & \multicolumn{2}{c|}{Earth} & \multicolumn{2}{c|}{Mercury} & \multicolumn{4}{c|}{Corona} & \multicolumn{3}{c|}{GCS} & Comment\\
          & $t_e$ & $v_e$ & $t_m$ & $v_{me}$ &No. & $t_c$ & Width & Direction & $t_{GCS}$ & $v_{GCS}$ & Direction & \\
          &  UT & km s$^{-1}$ & UT & km s$^{-1}$ & & UT & deg & & UT & km s$^{-1}$ & & \\
 \hline
 %          &          &     &          &     &    & 11 19:24 & $>271$ & To STA-     & && & Backside \\
         E0 & 16 04:05 & $380$ & 13 09:00 & $400$ & C0 & 12 06:00 & halo   & To Earth    & 10:20 & $641$ & W03S02 & \\
            &          &       &          &       &    & 12 13:25 & $124$  & To Earth+,N &       &       &        & Out of ecliptic plane\\
         E1 & 17 19:00 & $350$ & 13 20:50 & $290$ & C1 & 12 16:36 & halo   & To Earth+?  & 20:15?& $827$?& W31N09?& Flank? \\
            &          &       &          &       &    & 12 23:06 & halo   & To STA-     &       &       &        & Backside \\
            &          &       &          &       &    & 13 16:36 & $104$  & To STA+,S   &       &       &        & Backside \\
            &          &       &          &       &    & 14 08:48 & halo   & To STA      &       &       &        & Backside \\
         E2 & 18 16:10 & $400$ & 15 20:20 & $400$ & C2 & 14 11:42 & ?      & ?           &       &       &        & Stealth?\\
            &          &       &          &       &    & 14 16:00 & $145$  & To STB+,N   &       &       &        & Backside \\
            &          &       &          &       & L1 & 15 02:24 & $138$  & To Earth+   & 10:35 & $397$ & W46S05 & Limb \\
            &          &       &          &       &    & 15 09:48 & $112$  & To STB,S    &       &       &        & Backside \\
         E3 & 19 13:45 & $500$ & 17 07:05 & $490$ & C3 & 16 10:00 & halo   & To Earth    & 13:20 & $858$ & W02N00 & \\
            &          &       &          &       &    & 16 12:48 & $243$  & To STB+     &       &       &        & Backside \\
         E4 & 21 02:30 & $500$ & 18 21:05 & $500$ & C4 & 17 03:48 & $179$  & To Earth    & 06:55 & $857$ & W04S08 & \\
            &          &       &          &       &    & 17 05:12 & $121$  & To STA-     &       &       &        & Backside \\
     E4$^*$ &      &       &          &       & C5 & 18 01:36 & halo   & To Earth-   & 04:30 & $1075$& E35S09 & Flank \\
            &          &       &          &       &    & 18 23:24 & $133$  & To STA-,N   &       &       &        & Backside \\
 \hline
\end{tabular}
        $t_e$ and $t_m$ are the arrival times of the ejecta at Earth and Mercury, respectively. $v_e$ is the in-situ speed of
        the ejecta and $v_{me}$ is the transit speed of the ejecta from Mercury to Earth.
        $t_c$ is the first appearance in the field of view of the SOHO/LASCO, and
        the `Width' is the apparent angular width. `halo' means the angular width is $360^\circ$. The two parameters $t_c$ and `Width' are adopted from the LASCO CME catalog\cite{Yashiro_etal_2004}.
        The `Direction' under the column `Corona' is estimated by combining the images from the SOHO/LASCO and STEREO/COR2s. `STA' and `STB' stand for the twin STEREO spacecraft A and B. The `+' sign
        means that the direction of the CME is west to the Sun-observer line. `S' or `N' means that the CME's propagation direction
        is not near the ecliptic plane but toward the high latitude beneath or above the ecliptic plane. The question marks mean that the CME's parameters are not clear due to contamination by other CMEs. Seven potentially
        Earth-encountered CMEs are labeled as `C0' through `C5' and `L1' in the column `No.'.
        The columns of `GCS' list the parameters of the CMEs at 20 solar radii obtained by the GCS model, including the
        time, $t_{GCS}$, the real speed, $v_{GCS}$ and the propagation direction viewed from Earth. In the last column, `Limb' means
        that the CME is more than $45^\circ$ apart from the Sun-Earth line, and `Flank' means that the CME is still able to sweep through the
        Earth with its flank. The event marked as `E4$^*$' means that CME `C5' may catch up with the preceding one `C4' and form
        a complex ejecta as E4 at Earth.
\end{center}
\end{table*}

\newpage
\newpage

\begin{figure*}[h]
\begin{center}
\includegraphics[width=\hsize]{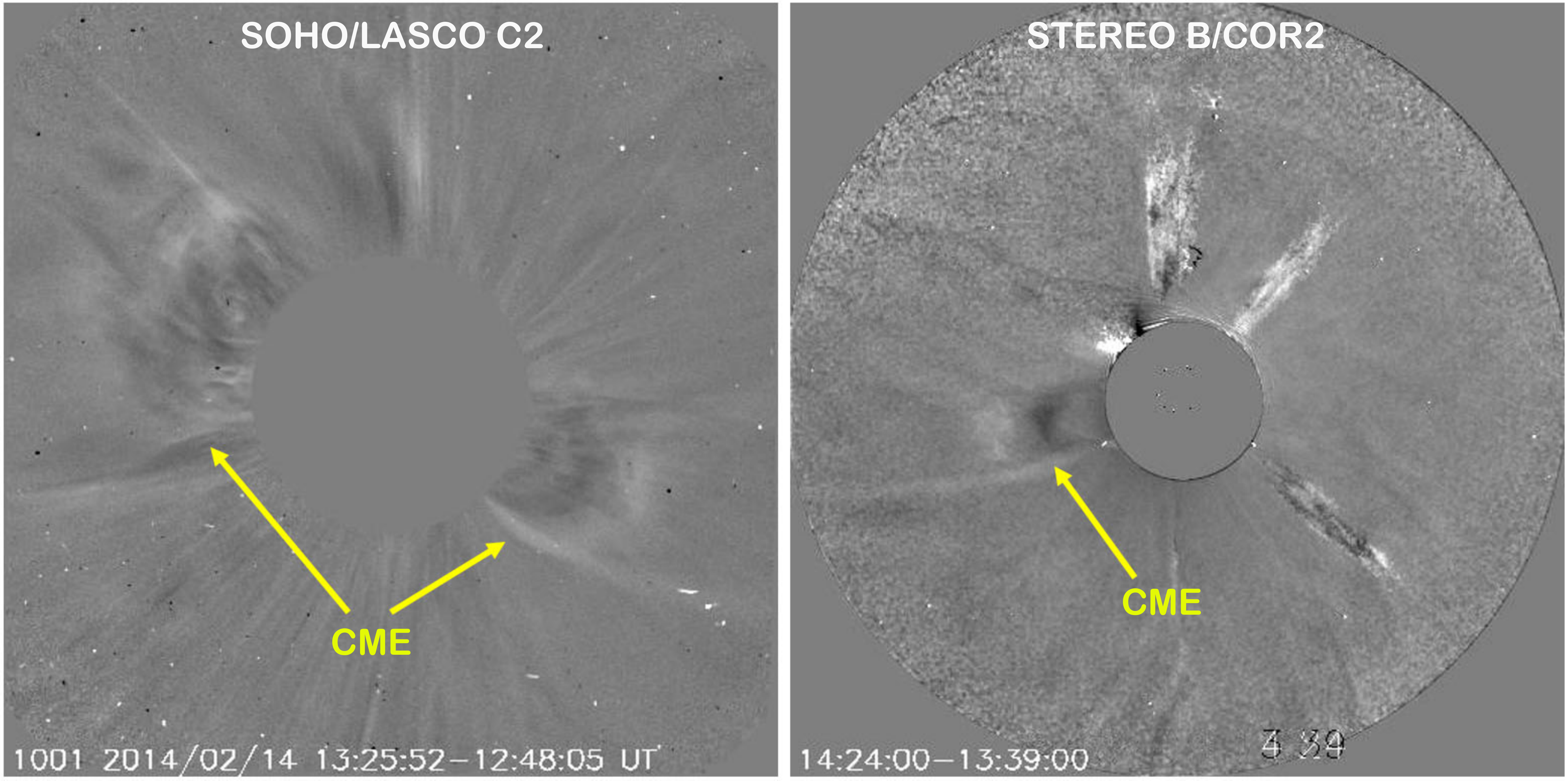}
	\caption{Most possible CME candidate for the magnetic cloud in the coronagraphs. Panel {\bf a}: The difference image 
	taken by the SOHO/LASCO C2 camera. Panel {\bf b}: The difference image taken by the STEREO-B/COR2 camera. The candidate 
	CME structures are denoted by the arrows.}\label{fig:cme}
\end{center}
\end{figure*}
\section{Identify the corresponding CME and source region of the magnetic cloud of interest `E2'}\label{supp:mc_source}
According to the above DIPS model result, the magnetic cloud speed is about $400$ km s$^{-1}$, and the expected onset time
of the corresponding CME is at about 09:00 UT on February 14. We check all the CMEs with apparent angular width larger than $90^\circ$
during February 13 -- 15, which can be found in Table~\ref{tb:associations}. There are 6 CMEs for consideration, among which
four CMEs were almost backside as identified in the previous subsection. CME `L1' was a limb event and too slow to be the
corresponding CME of the magnetic cloud. CME `C2' is not in the LASCO CME catalog. By manually checking the coronagraphs images,
we find there was a weak CME entering the field of view of
SOHO LASCO at about 11:42 UT on February 14, right behind the strong CME appearing on 08:48 UT.
Two snapshots taken by SOHO/LASCO C2 and STEREO-B/COR2 cameras, respectively,
are shown in Figure~\ref{fig:cme}. We do not show the images from STEREO-A, because
the quality is not good enough. In the left panel, there were three CME-like structures, one toward the south-west
in the plane-of-the-sky and the other two, very close to each other, toward the north-east. The upper one in the
north-east direction can be identified as a high-latitude CME toward the east of the Sun-Earth line from the SOHO/LASCO
and STEREO-A and B's COR2 images (not shown here). However, it is not clear whether or not the lower one belonged to the
same CME of the south-west one. If it was true, the CME right faced on Earth as expected. However, in
the right panel of the figure, we can only recognize one CME structure toward the east from the view of STEREO-B, which
corresponds to the south-west structure in the SOHO/LASCO image. This makes the identification ambiguous.

Even if the south-west CME was the most probable candidate, the EUV images taken by the Atmospheric Imaging
Assembly~\citep{Lemen_etal_2012} onboard the Solar Dynamics Observatory (SDO) show no signature of the CME on the solar
surface in a reasonable period before the CME
appeared in the field of view of the SOHO/LASCO. Thus, it is also possible that the magnetic cloud observed by MESSENGER
corresponds to a stealth CME~\citep[e.g.,][]{Robbrecht_etal_2009, Ma_etal_2010, Wang_etal_2011, HowardT_Harrison_2013}.

\begin{figure*}[h]
\begin{center}
\includegraphics[width=0.85\hsize]{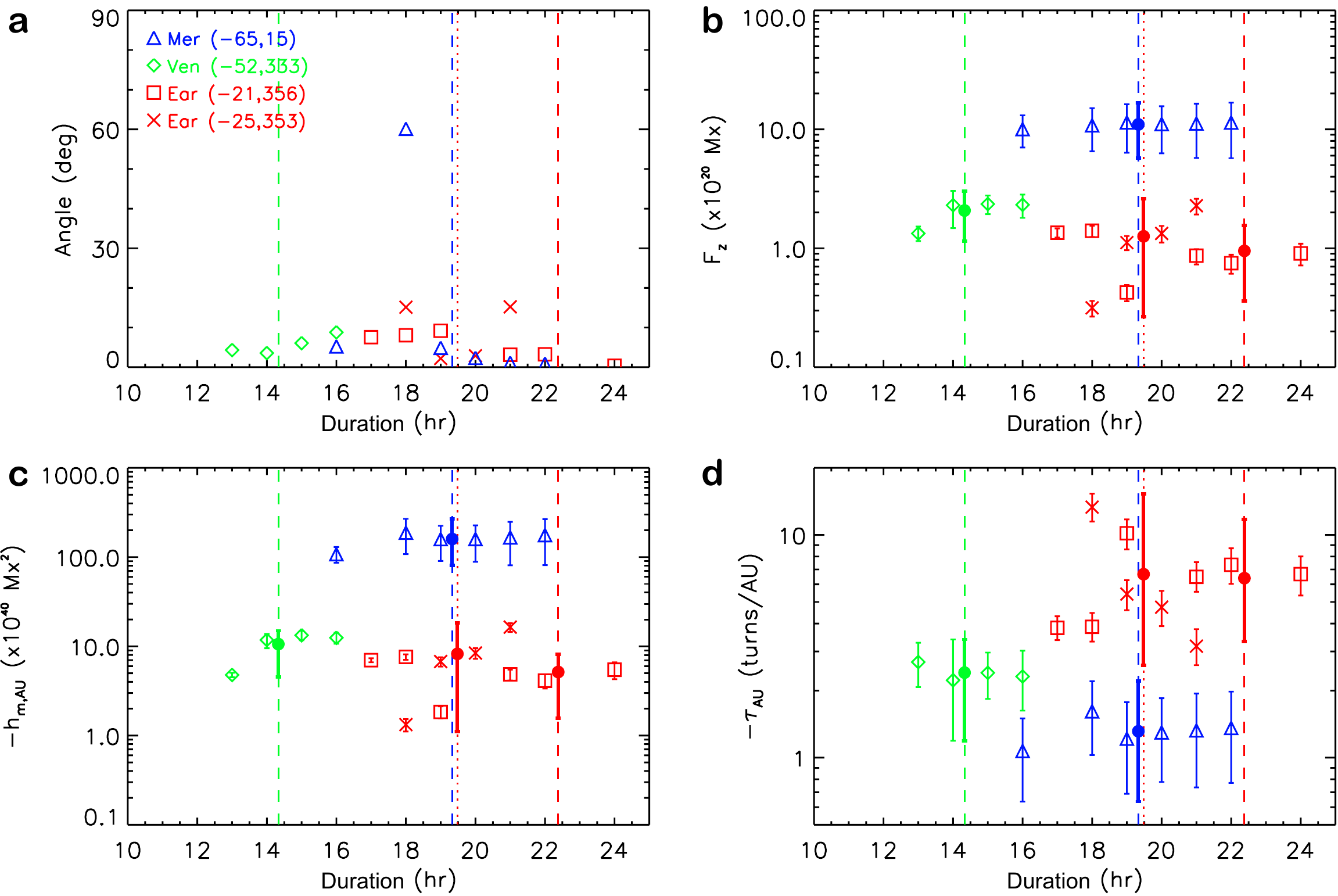}
	\caption{Same as Fig.\ref{fig:fitting} except that the fitting results of the magnetic cloud at the Earth by changing the 
	uniform compression ratio to the linearly-decreasing compression ratio are added as $\times$ symbols for comparison. The red dotted vertical line 
indicates the duration of the magnetic cloud by using the new compression ratio.}\label{fig:fitting_cratio}
\end{center}
\end{figure*}

\section{Influence of the non-expansion assumption on the fitting results}\label{supp:expansion}
\begin{figure}[t]
\begin{center}
\includegraphics[width=\hsize]{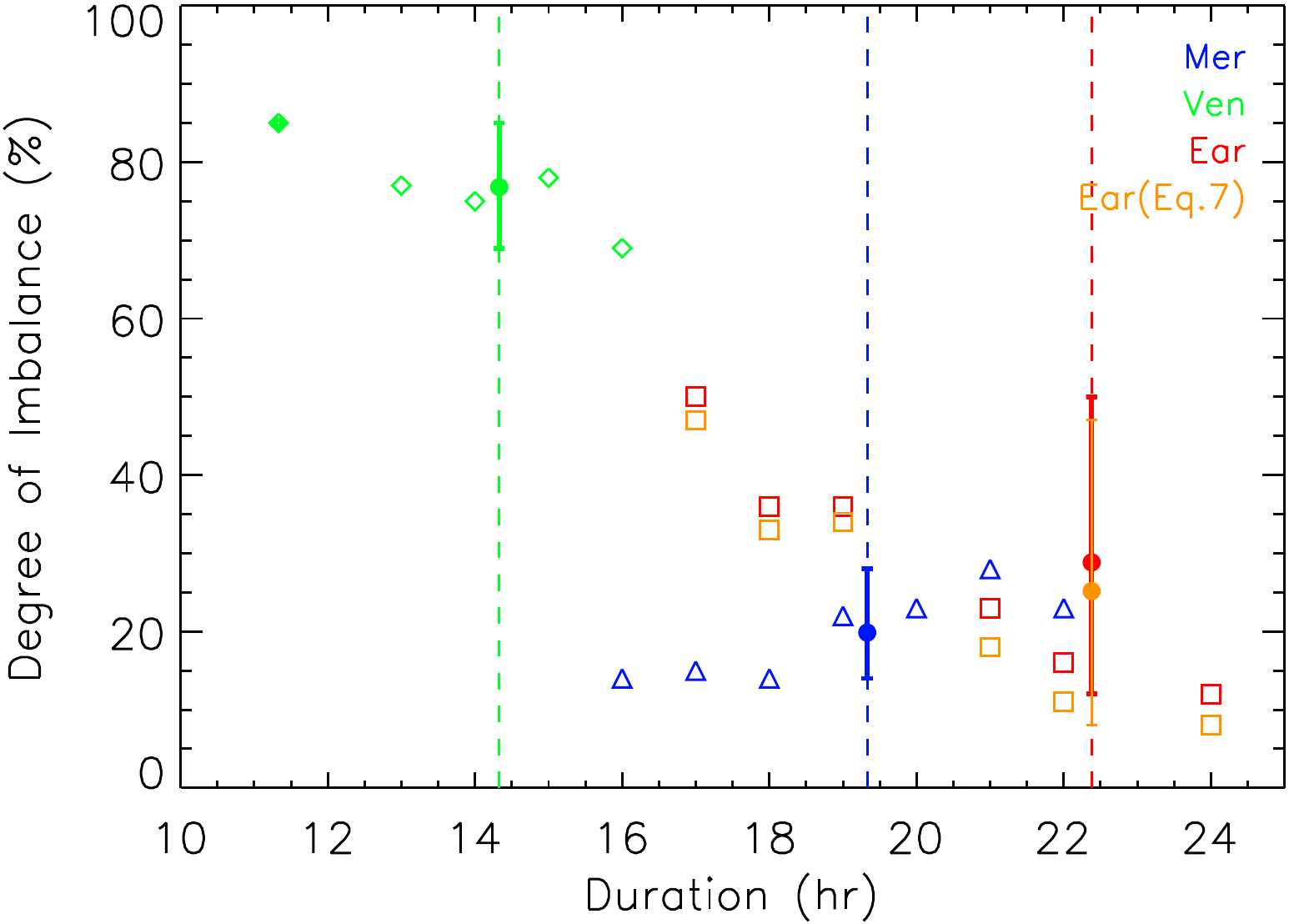}
\caption{Same as Fig.\ref{fig:imbalance-duration} except that we set an expansion speed to be $20$ km s$^{-1}$ when fitting the magnetic cloud at Mercury.}\label{fig:imbalance-duration_test}
\end{center}
\end{figure}

\begin{figure*}[h]
\begin{center}
\includegraphics[width=0.85\hsize]{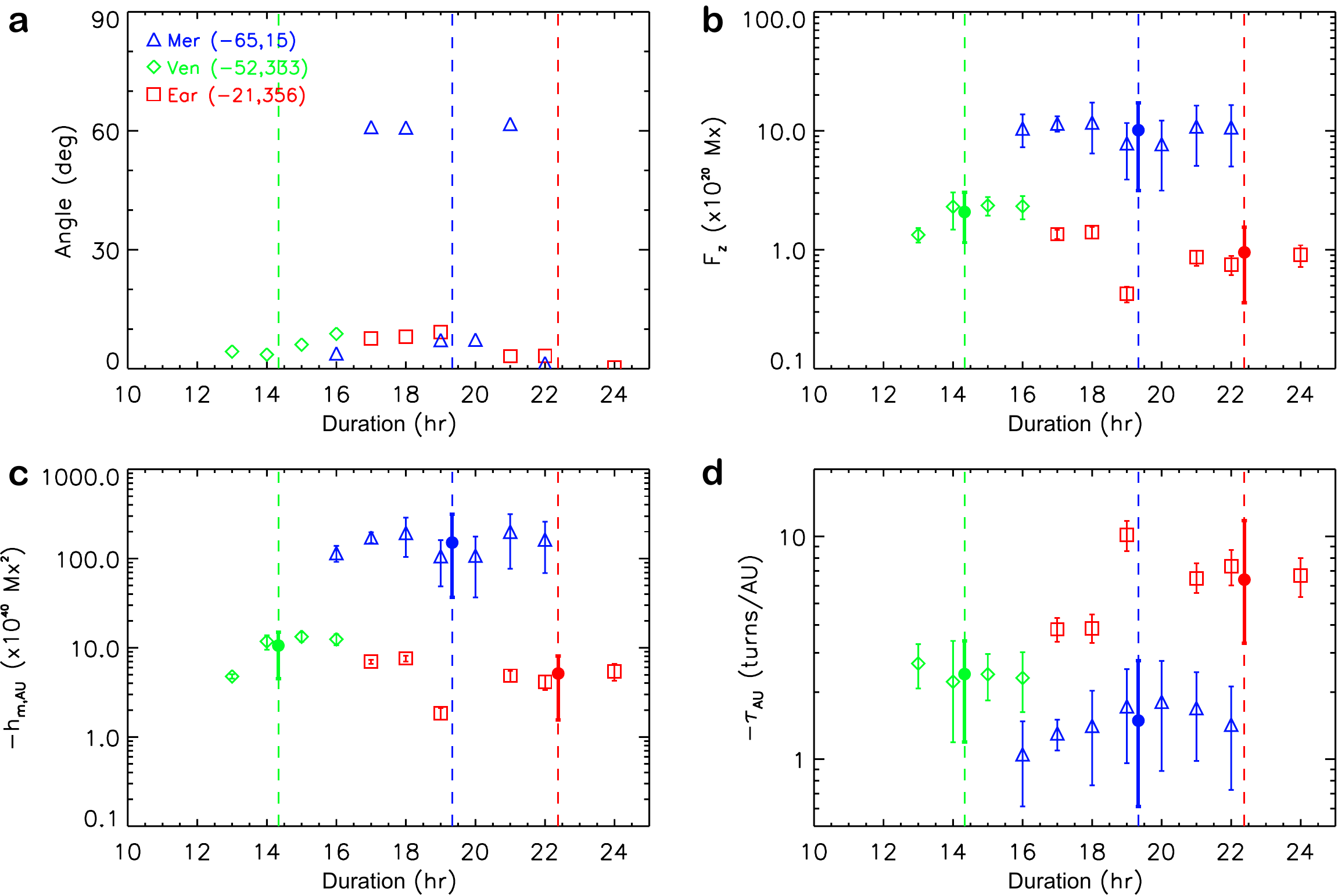}
\caption{Same as Fig.\ref{fig:fitting} except that we set an expansion speed to be $20$ km s$^{-1}$ when fitting the magnetic cloud at Mercury.}\label{fig:fitting_test}
\end{center}
\end{figure*}

The Wind data suggest that the magnetic cloud might experience a weak expansion with a speed of about 20 km s$^{-1}$at 1 AU (see $v_x$ profile in Fig.\ref{fig:shock_recover}).
However, in our fitting procedure for the magnetic cloud at Mercury and Venus, the expansion speed is assumed to be zero, which 
might influence the fitting results. To test how significant the influence will be, we set the expansion speed to be $20$ km s$^{-1}$ for the cloud at Mercury and run the fitting code again. 
The test results are shown in Figures~\ref{fig:fitting_test} and \ref{fig:imbalance-duration_test}, which correspond to Figures~\ref{fig:fitting} and \ref{fig:imbalance-duration} in the main text, respectively. 
By comparing the blue symbols in the two sets of figures, it could be found that there is no evident difference in the axial flux, magnetic helicity, twist and the degree of imbalance, 
except for two more test cases with orientations deviating largely from the final orientation determined for non-expansion cases.
The comparison suggests that the assumption of non-expansion speed has small influence on our results and conclusions.

\bibliographystyle{agu}
\bibliography{../../../ahareference}

\begin{thebibliography}{95}
\providecommand{\natexlab}[1]{#1}
\expandafter\ifx\csname urlstyle\endcsname\relax
  \providecommand{\doi}[1]{doi:\discretionary{}{}{}#1}\else
  \providecommand{\doi}{doi:\discretionary{}{}{}\begingroup
  \urlstyle{rm}\Url}\fi

\bibitem[{\textit{Anderson et~al.}(2007)\textit{Anderson, Acu\~{n}a, Lohr,
  Scheifele, Raval, Korth, and Slavin}}]{Anderson_etal_2007}
Anderson, B.~J., M.~H. Acu\~{n}a, D.~A. Lohr, J.~Scheifele, A.~Raval, H.~Korth,
  and J.~A. Slavin, The magnetometer instrument on {MESSENGER}, \textit{Space
  Sci. Rev.}, \textit{131}, 417--450, 2007.

\bibitem[{\textit{Antiochos et~al.}(1999)\textit{Antiochos, DeVore, and
  Klimchuk}}]{Antiochos_etal_1999}
Antiochos, S.~K., C.~R. DeVore, and J.~A. Klimchuk, A model for solar coronal
  mass ejections, \textit{Astrophys. J.}, \textit{510}, 485--493, 1999.

\bibitem[{\textit{Aulanier et~al.}(2010)\textit{Aulanier, T{\"o}r{\"o}k,
  D\'emoulin, and DeLuca}}]{Aulanier_etal_2010}
Aulanier, G., T.~T{\"o}r{\"o}k, P.~D\'emoulin, and E.~E. DeLuca, Formation of
  torus-unstable flux ropes and electric currents in erupting sigmoids,
  \textit{Astrophys. J.}, \textit{708}, 314--333, 2010.

\bibitem[{\textit{Aulanier et~al.}(2012)\textit{Aulanier, Janvier, and
  Schmieder}}]{Aulanier_etal_2012}
Aulanier, G., M.~Janvier, and B.~Schmieder, The standard flare model in three
  dimensions. {I}. strong-to-weak shear transition in post-flare loops,
  \textit{Astron. \& Astrophys.}, \textit{543}, A110(14pp), 2012.

\bibitem[{\textit{Brueckner et~al.}(1995)}]{Brueckner_etal_1995}
Brueckner, G.~E., et~al., The large angle spectroscopic coronagraph ({LASCO}),
  \textit{Sol. Phys.}, \textit{162}, 357--402, 1995.

\bibitem[{\textit{Burlaga et~al.}(1981)\textit{Burlaga, Sittler, Mariani, and
  Schwenn}}]{Burlaga_etal_1981}
Burlaga, L., E.~Sittler, F.~Mariani, and R.~Schwenn, Magnetic loop behind an
  interplanetary shock: {Voyager}, {Helios}, and {IMP} 8 observations,
  \textit{J. Geophys. Res.: Space Phys.}, \textit{86}, 6673--6684, 1981.

\bibitem[{\textit{Cane}(2000)}]{Cane_2000}
Cane, H.~V., Coronal mass ejections and forbush decreases, \textit{Space Sci.
  Rev.}, \textit{93}, 55--77, 2000.

\bibitem[{\textit{Chintzoglou et~al.}(2015)\textit{Chintzoglou, Patsourakos,
  and Vourlidas}}]{Chintzoglou_etal_2015}
Chintzoglou, G., S.~Patsourakos, and A.~Vourlidas, Formation of magnetic flux
  ropes during confined flaring well before the onset of a pair of major
  coronal mass ejections, \textit{Astrophys. J.}, \textit{809}, 34(18pp), 2015.

\bibitem[{\textit{Cid et~al.}(2002)\textit{Cid, Hidalgo, Nieves-Chinchilla,
  Sequeiros, and Vi\~{n}as}}]{Cid_etal_2002}
Cid, C., M.~A. Hidalgo, T.~Nieves-Chinchilla, J.~Sequeiros, and A.~F.
  Vi\~{n}as, Plasma and magnetic field inside magnetic clouds: a global study,
  \textit{Sol. Phys.}, \textit{207}, 187--198, 2002.

\bibitem[{\textit{Crooker and Intriligator}(1996)}]{Crooker_Intriligator_1996}
Crooker, N.~V., and D.~S. Intriligator, A magnetic cloud as a distended flux
  rope occlusion in the heliospheric current sheet, \textit{J. Geophys. Res.:
  Space Phys.}, \textit{101}, 24,343--24,348, 1996.

\bibitem[{\textit{Dasso et~al.}(2006)\textit{Dasso, Mandrini, D\'{e}moulin, and
  Luoni}}]{Dasso_etal_2006}
Dasso, S., C.~H. Mandrini, P.~D\'{e}moulin, and M.~L. Luoni, A new
  model-independent method to compute magnetic helicity in magnetic clouds,
  \textit{Astron. \& Astrophys.}, \textit{455}, 349--359, 2006.

\bibitem[{\textit{Daughton et~al.}(2011)\textit{Daughton, Roytershteyn,
  Karimabadi, Yin, Albright, Bergen, and Bowers}}]{Daughton_etal_2011}
Daughton, W., V.~Roytershteyn, H.~Karimabadi, L.~Yin, B.~J. Albright,
  B.~Bergen, and K.~J. Bowers, Role of electron physics in the development of
  turbulent magnetic reconnection in collisionless plasmas, \textit{Nature
  Phys.}, \textit{7}, 539--542, 2011.

\bibitem[{\textit{D\'{e}moulin and Dasso}(2009)}]{Demoulin_Dasso_2009}
D\'{e}moulin, P., and S.~Dasso, Causes and consequences of magnetic cloud
  expansion, \textit{Astron. \& Astrophys.}, \textit{498}, 551--566, 2009.

\bibitem[{\textit{Du et~al.}(2007)\textit{Du, Wang, and Hu}}]{Du_etal_2007}
Du, D., C.~Wang, and Q.~Hu, Propagation and evolution of a magnetic cloud from
  {ACE} to {Ulysses}, \textit{J. Geophys. Res.: Space Phys.}, \textit{112},
  A09,101, 2007.

\bibitem[{\textit{Dungey and Loughhead}(1954)}]{Dungey_Loughhead_1954}
Dungey, J.~W., and R.~E. Loughhead, Twisted magnetic fields in conducting
  fluids, \textit{Aust. J. Phys.}, \textit{7}, 5--13, 1954.

\bibitem[{\textit{Gary and Moore}(2004)}]{Gary_Moore_2004}
Gary, G.~A., and R.~L. Moore, Eruption of a multi-turn helical magnetic flux
  tube in a large flare: Evidence for external and internal reconnection that
  fits the breakout model of solar magnetic eruptions, \textit{Astrophys. J.},
  \textit{611}, 545--556, 2004.

\bibitem[{\textit{Goldstein}(1983)}]{Goldstein_1983}
Goldstein, H., On the field configuration in magnetic clouds, in \textit{Sol.
  Wind Five}, p. 731, NASA Conf. Publ. 2280, Washington D. C., 1983.

\bibitem[{\textit{G{\'o}mez et~al.}(2008)\textit{G{\'o}mez, Marscher, Jorstad,
  Agudo, and Roca-Sogorb}}]{Gomez_etal_2008}
G{\'o}mez, J.~L., A.~P. Marscher, S.~G. Jorstad, I.~Agudo, and M.~Roca-Sogorb,
  Faraday rotation and polarization gradients in the jet of {3C} 120:
  Interaction with the external medium and a helical magnetic field?,
  \textit{Astrophys. J.}, \textit{681}, L69--L72, 2008.

\bibitem[{\textit{Good et~al.}(2015)\textit{Good, Forsyth, Raines, Gershman,
  Slavin, and Zurbuchen}}]{Good_etal_2015}
Good, S., R.~Forsyth, J.~M. Raines, D.~J. Gershman, J.~A. Slavin, and T.~H.
  Zurbuchen, Radial evolution of a magnetic cloud: {MESSENGER}, {STEREO}, and
  {Venus} {Express} observations, \textit{Astrophys. J.}, \textit{807},
  177(12pp), 2015.

\bibitem[{\textit{Gopalswamy et~al.}(2000)\textit{Gopalswamy, Lara, Lepping,
  Kaiser, Berdichevsky, and {St. Cyr}}}]{Gopalswamy_etal_2000}
Gopalswamy, N., A.~Lara, R.~P. Lepping, M.~L. Kaiser, D.~Berdichevsky, and
  O.~C. {St. Cyr}, Interplanetary acceleration of coronal mass ejections,
  \textit{Geophys. Res. Lett.}, \textit{27}, 145--148, 2000.

\bibitem[{\textit{Gosling}(2012)}]{Gosling_2012}
Gosling, J.~T., Magnetic reconnection in the solar wind, \textit{Space Sci.
  Rev.}, \textit{172}, 187--200, 2012.

\bibitem[{\textit{Grotzinger et~al.}(2012)}]{Grotzinger_etal_2012}
Grotzinger, J.~P., et~al., Mars science laboratory mission and science
  investigation, \textit{Space Sci. Rev.}, \textit{170}, 5--56, 2012.

\bibitem[{\textit{Guo et~al.}(2018)\textit{Guo, Lillis, Wimmer-Schweingruber,
  and {et al.}}}]{GuoJ_etal_2018}
Guo, J., R.~Lillis, R.~Wimmer-Schweingruber, and {et al.}, Measurements of
  {Forbush} decreases at {Mars}: both by {MSL} on ground and by {MAVEN} in
  orbit, \textit{Astron. \& Astrophys.}, \textit{in press},
  DOI:/10.1051/0004--6361/201732,087, 2018.

\bibitem[{\textit{Guo et~al.}(2017)}]{GuoJ_etal_2017}
Guo, J., et~al., Dependence of the martian radiation environment on atmospheric
  depth: Modeling and measurement, \textit{J. Geophys. Res.: Planets},
  \textit{122}, 329--341, 2017.

\bibitem[{\textit{Hassler et~al.}(2012)}]{Hassler_etal_2012}
Hassler, D.~M., et~al., The {Radiation Assessment Detector} ({RAD})
  investigation, \textit{Space Sci. Rev.}, \textit{170}, 503--558, 2012.

\bibitem[{\textit{Hidalgo et~al.}(2002)\textit{Hidalgo, Cid, Vinas, and
  Sequeiros}}]{Hidalgo_etal_2002}
Hidalgo, M.~A., C.~Cid, A.~F. Vinas, and J.~Sequeiros, A non-force-free
  approach to the topology of magnetic clouds in the solar wind, \textit{J.
  Geophys. Res.: Space Phys.}, \textit{107}, 1002, 2002.

\bibitem[{\textit{Hood and Priest}(1981)}]{Hood_Priest_1981}
Hood, A.~W., and E.~R. Priest, Critical conditions for magnetic instabilities
  in force-free coronal loops, \textit{Geophys. and Astrophys. Fluid Dynamics},
  \textit{17}, 297--318, 1981.

\bibitem[{\textit{Howard et~al.}(2008)\textit{Howard, Moses, Vourlidas, and {et
  al.}}}]{Howard_etal_2008}
Howard, R.~A., J.~Moses, A.~Vourlidas, and {et al.}, Sun earth connection
  coronal and heliospheric investigation {(SECCHI)}, \textit{Space Sci. Rev.},
  \textit{136}, 67--115, 2008.

\bibitem[{\textit{Howard and Harrison}(2013)}]{HowardT_Harrison_2013}
Howard, T.~A., and R.~A. Harrison, Stealth coronal mass ejections: A
  perspective, \textit{Sol. Phys.}, \textit{285}, 269--280, 2013.

\bibitem[{\textit{Hu and Sonnerup}(2002)}]{Hu_Sonnerup_2002}
Hu, Q., and B.~U.~O. Sonnerup, Reconstruction of magnetic clouds in the solar
  wind: Orientations and configurations, \textit{J. Geophys. Res.: Space
  Phys.}, \textit{107}, 1142, 2002.

\bibitem[{\textit{Hu et~al.}(2015)\textit{Hu, Qiu, and Krucker}}]{Hu_etal_2015}
Hu, Q., J.~Qiu, and S.~Krucker, Magnetic field line lengths inside
  interplanetary magnetic flux ropes, \textit{J. Geophys. Res.: Space Phys.},
  \textit{120}, 5266--5283, 2015.

\bibitem[{\textit{Karpen et~al.}(2012)\textit{Karpen, Antiochos, and
  DeVore}}]{Karpen_etal_2012}
Karpen, J.~T., S.~K. Antiochos, and C.~R. DeVore, The mechanisms for the onset
  and explosive eruption of coronal mass ejections and eruptive flares,
  \textit{Astrophys. J.}, \textit{760}, 81, 2012.

\bibitem[{\textit{Kopp and Pneuman}(1976)}]{Kopp_Pneuman_1976}
Kopp, R.~A., and G.~W. Pneuman, Magnetic reconnection in the corona and the
  loop prominence phenomenon, \textit{Sol. Phys.}, \textit{50}, 85--98, 1976.

\bibitem[{\textit{Kruskal et~al.}(1958)\textit{Kruskal, Johnson, Gottlieb, and
  Goldman}}]{Kruskal_etal_1958}
Kruskal, M.~D., J.~L. Johnson, M.~B. Gottlieb, and L.~M. Goldman, Hydromagnetic
  instability in a stellarator, \textit{Phys. Fluids}, \textit{1}, 421--429,
  1958.

\bibitem[{\textit{Leitner et~al.}(2007)\textit{Leitner, Farrugia, M{o"}stl,
  Ogilvie, Galvin, Schwenn, and Biernat}}]{Leitner_etal_2007}
Leitner, M., C.~J. Farrugia, C.~M{o"}stl, K.~W. Ogilvie, A.~B. Galvin,
  R.~Schwenn, and H.~K. Biernat, Consequences of the force-free model of
  magnetic clouds for their heliospheric evolution, \textit{J. Geophys. Res.:
  Space Phys.}, \textit{112}, A06,113, 2007.

\bibitem[{\textit{Lemen et~al.}(2012)}]{Lemen_etal_2012}
Lemen, J.~R., et~al., The atmospheric imaging assembly ({AIA}) on the solar
  dynamics observatory ({SDO}), \textit{Sol. Phys.}, \textit{275}, 17--40,
  2012.

\bibitem[{\textit{Lepping et~al.}(1990)\textit{Lepping, Jones, and
  Burlaga}}]{Lepping_etal_1990}
Lepping, R.~P., J.~A. Jones, and L.~F. Burlaga, Magnetic field structure of
  interplanetary magnetic clouds at 1 {AU}, \textit{J. Geophys. Res.: Space
  Phys.}, \textit{95}, 11,957--11,965, 1990.

\bibitem[{\textit{Lepping et~al.}(2006)\textit{Lepping, Berdichevsky, Wu,
  Szabo, Narock, Mariani, Lazarus, and Quivers}}]{Lepping_etal_2006}
Lepping, R.~P., D.~B. Berdichevsky, C.-C. Wu, A.~Szabo, T.~Narock, F.~Mariani,
  A.~J. Lazarus, and A.~J. Quivers, A summary of {WIND} magnetic clouds for
  years 1995--2003: model-fitted parameters, associated errors and
  classifications, \textit{Ann. Geophys.}, \textit{24}, 215--245, 2006.

\bibitem[{\textit{Lepping et~al.}(1995)}]{Lepping_etal_1995}
Lepping, R.~P., et~al., The {Wind} magnetic field investigation, \textit{Space
  Sci. Rev.}, \textit{71}, 207--229, 1995.

\bibitem[{\textit{Liu et~al.}(2016{\natexlab{a}})\textit{Liu, Wang, Wang, Shen,
  Ye, Liu, Chen, Zhang, and Wang}}]{LiuL_etal_2016}
Liu, L., Y.~Wang, J.~Wang, C.~Shen, P.~Ye, R.~Liu, J.~Chen, Q.~Zhang, and
  S.~Wang, Why is a flare-rich active region {CME}-poor?, \textit{Astrophys.
  J.}, \textit{826}, 119(10pp), 2016{\natexlab{a}}.

\bibitem[{\textit{Liu et~al.}(2016{\natexlab{b}})\textit{Liu, Kliem, Titov,
  Chen, Wang, Wang, Liu, Xu, and Wiegelmann}}]{LiuR_etal_2016}
Liu, R., B.~Kliem, V.~S. Titov, J.~Chen, Y.~Wang, H.~Wang, C.~Liu, Y.~Xu, and
  T.~Wiegelmann, Structure, stability, and evolution of magnetic flux ropes
  from the perspective of magnetic twist, \textit{Astrophys. J.}, \textit{818},
  148(22pp), 2016{\natexlab{b}}.

\bibitem[{\textit{Longcope and Beveridge}(2007)}]{Longcope_Beveridge_2007}
Longcope, D.~W., and C.~Beveridge, A quantitative, topological model of
  reconnection and flux rope formation in a two-ribbon flare,
  \textit{Astrophys. J.}, \textit{669}, 621--635, 2007.

\bibitem[{\textit{Lundquist}(1950)}]{Lundquist_1950}
Lundquist, S., Magnetohydrostatic fields, \textit{Ark. Fys.}, \textit{2}, 361,
  1950.

\bibitem[{\textit{Ma et~al.}(2010)\textit{Ma, Attrill, Golub, and
  Lin}}]{Ma_etal_2010}
Ma, S., G.~D.~R. Attrill, L.~Golub, and J.~Lin, Statistical study of coronal
  mass ejections with and without distinct low coronal signatures,
  \textit{Astrophys. J.}, \textit{722}, 289--301, 2010.

\bibitem[{\textit{Manchester et~al.}(2004)\textit{Manchester, Gombosi, Roussev,
  {De Zeeuw}, Sokolov, Powell, Toth, and Opher}}]{Manchester_etal_2004a}
Manchester, W.~B., T.~I. Gombosi, I.~Roussev, D.~L. {De Zeeuw}, I.~V. Sokolov,
  K.~G. Powell, G.~Toth, and M.~Opher, Three-dimensional mhd simulation of a
  flux rope driven {CME}, \textit{J. Geophys. Res.: Space Phys.},
  \textit{109(A1)}, A01,102, 2004.

\bibitem[{\textit{Manchester et~al.}(2014)\textit{Manchester, Kozyra, Lepri,
  and Lavraud}}]{Manchester_etal_2014}
Manchester, W.~B., J.~U. Kozyra, S.~T. Lepri, and B.~Lavraud, Simulation of
  magnetic cloud erosion during propagation, \textit{J. Geophys. Res.: Space
  Phys.}, \textit{119}, 5449--5464, 2014.

\bibitem[{\textit{Marscher et~al.}(2008)}]{Marscher_etal_2008}
Marscher, A.~P., et~al., The inner jet of an active galactic nucleus as
  revealed by a radio-to-$\gamma$-ray outburst, \textit{Nature}, \textit{452},
  966--969, 2008.

\bibitem[{\textit{Marubashi}(1986)}]{Marubashi_1986}
Marubashi, K., Structure of the interplanetary magnetic clouds and their solar
  origins, \textit{Adv. in Space Res.}, \textit{6}, 335--338, 1986.

\bibitem[{\textit{Moore et~al.}(2001)\textit{Moore, Sterling, Hudson, and
  Lemen}}]{Moore_etal_2001}
Moore, R.~L., A.~C. Sterling, H.~S. Hudson, and J.~R. Lemen, Onset of the
  magnetic explosion in solar flares and coronal mass ejections,
  \textit{Astrophys. J.}, \textit{552}, 833--848, 2001.

\bibitem[{\textit{M{\"o}stl and Davies}(2013)}]{Mostl_Davies_2013}
M{\"o}stl, C., and J.~A. Davies, Speeds and arrival times of solar transients
  approximated by self-similar expanding circular fronts, \textit{Sol. Phys.},
  \textit{285}, 411--423, 2013.

\bibitem[{\textit{Mulligan and Russell}(2001)}]{Mulligan_Russell_2001}
Mulligan, T., and C.~T. Russell, Multispacecraft modeling of the flux rope
  structure of interplanetary coronal mass ejections: Cylindrically symmetric
  versus nonsymmetric topologies, \textit{J. Geophys. Res.: Space Phys.},
  \textit{106(A6)}, 10,581--10,596, 2001.

\bibitem[{\textit{Mulligan et~al.}(2001)\textit{Mulligan, Russell, Anderson,
  and Acuna}}]{Mulligan_etal_2001}
Mulligan, T., C.~T. Russell, B.~J. Anderson, and M.~H. Acuna, Multiple
  spacecraft flux rope modeling of the {Bastille Day} magnetic cloud,
  \textit{Geophys. Res. Lett.}, \textit{28(23)}, 4417--4420, 2001.

\bibitem[{\textit{Myers et~al.}(2015)\textit{Myers, Yamada, Ji, Yoo, Fox,
  Jara-Almonte, Savcheva, and Deluca}}]{Myers_etal_2015}
Myers, C.~E., M.~Yamada, H.~Ji, J.~Yoo, W.~Fox, J.~Jara-Almonte, A.~Savcheva,
  and E.~E. Deluca, A dynamic magnetic tension force as the cause of failed
  solar eruptions, \textit{Nature}, \textit{528}, 526--529, 2015.

\bibitem[{\textit{Nakwacki et~al.}(2011)\textit{Nakwacki, Dasso, D{\'e}moulin,
  Mandrini, and Gulisano}}]{Nakwacki_etal_2011}
Nakwacki, M.~S., S.~Dasso, P.~D{\'e}moulin, C.~H. Mandrini, and A.~M. Gulisano,
  Dynamical evolution of a magnetic cloud from the {Sun} to 5.4 {AU},
  \textit{Astron. \& Astrophys.}, \textit{535}, A52, 2011.

\bibitem[{\textit{Nieves-Chinchilla et~al.}(2012)\textit{Nieves-Chinchilla,
  Colaninno, Vourlidas, Szabo, Lepping, Boardsen, Anderson, and
  Korth}}]{Nieves-Chinchilla_etal_2012}
Nieves-Chinchilla, T., R.~Colaninno, A.~Vourlidas, A.~Szabo, R.~P. Lepping,
  S.~A. Boardsen, B.~J. Anderson, and H.~Korth, Remote and in situ observations
  of an unusual earth-directed coronal mass ejection from multiple viewpoints,
  \textit{J. Geophys. Res.: Space Phys.}, \textit{117}, A06,106, 2012.

\bibitem[{\textit{Ogilvie et~al.}(1995)}]{Ogilvie_etal_1995}
Ogilvie, K.~W., et~al., {SWE}, a comprehensive plasma instrument for the {Wind}
  spacecraft, \textit{Space Sci. Rev.}, \textit{71}, 55--77, 1995.

\bibitem[{\textit{Owen et~al.}(1989)\textit{Owen, Hardee, and
  Cornwell}}]{Owen_etal_1989}
Owen, F.~N., P.~E. Hardee, and T.~J. Cornwell, High-resolution, high dynamic
  range {VLA} images of the {M87} jet at 2 centimeters, \textit{Astrophys. J.},
  \textit{340}, 698--707, 1989.

\bibitem[{\textit{Perley et~al.}(1984)\textit{Perley, Bridle, and
  Willis}}]{Perley_etal_1984}
Perley, R.~A., A.~H. Bridle, and A.~G. Willis, High-resolution {VLA}
  observations of the radio jet in {NGC} 6251, \textit{Astrophys. J.},
  \textit{54}, 291--334, 1984.

\bibitem[{\textit{Priest and Longcope}(2017)}]{Priest_Longcope_2017}
Priest, E.~R., and D.~W. Longcope, Flux-rope twist in eruptive flares and
  {CMEs}: Due to zipper and main-phase reconnection, \textit{Sol. Phys.},
  \textit{292}, 25(31pp), 2017.

\bibitem[{\textit{Qiu et~al.}(2007)\textit{Qiu, Hu, Howard, and
  Yurchyshyn}}]{Qiu_etal_2007}
Qiu, J., Q.~Hu, T.~A. Howard, and V.~B. Yurchyshyn, On the magnetic flux budget
  in low-corona magnetic reconnection and interplanetary coronal mass
  ejections, \textit{Astrophys. J.}, \textit{659}, 758--772, 2007.

\bibitem[{\textit{Riley and Crooker}(2004)}]{Riley_Crooker_2004}
Riley, P., and N.~U. Crooker, Kinematic treatment of coronal mass ejection
  evolution in the solar wind, \textit{Astrophys. J.}, \textit{600},
  1035--1042, 2004.

\bibitem[{\textit{Riley et~al.}(2003)\textit{Riley, Linker, Mikic, Odstrcil,
  Zurbuchen, and Lario}}]{Riley_etal_2003}
Riley, P., J.~A. Linker, Z.~Mikic, D.~Odstrcil, T.~H. Zurbuchen, and R.~P.
  Lario, D. amd~Lepping, Using an {MHD} simulation to interpret the global
  context of a coronal mass ejection observed by two spacecraft, \textit{J.
  Geophys. Res.: Space Phys.}, \textit{108}, 1272, 2003.

\bibitem[{\textit{Riley et~al.}(2004)}]{Riley_etal_2004}
Riley, P., et~al., Fitting flux ropes to a global {MHD} solution: a comparison
  of techniques, \textit{J. Atmos. Solar-Terres. Phys.}, \textit{66},
  1321--1331, 2004.

\bibitem[{\textit{Robbrecht et~al.}(2009)\textit{Robbrecht, Patsourakos, and
  Vourlidas}}]{Robbrecht_etal_2009}
Robbrecht, E., S.~Patsourakos, and A.~Vourlidas, No trace left behind: {STEREO}
  observation of a coronal mass ejection without low coronal signatures,
  \textit{Astrophys. J.}, \textit{701}, 283--291, 2009.

\bibitem[{\textit{Ruffenach et~al.}(2012)}]{Ruffenach_etal_2012}
Ruffenach, A., et~al., Multispacecraft observation of magnetic cloud erosion
  bymagnetic reconnection during propagation, \textit{J. Geophys. Res.: Space
  Phys.}, \textit{117}, A09,101, 2012.

\bibitem[{\textit{Ruffenach et~al.}(2015)}]{Ruffenach_etal_2015}
Ruffenach, A., et~al., Statistical study of magnetic cloud erosion by magnetic
  reconnection, \textit{J. Geophys. Res.: Space Phys.}, \textit{120}, 43--60,
  2015.

\bibitem[{\textit{Russell and Mulligan}(2002)}]{Russell_Mulligan_2002}
Russell, C.~T., and T.~Mulligan, The true dimensions of interplanetary coronal
  mass ejections, \textit{Adv. Space Res.}, \textit{29}, 301--306, 2002.

\bibitem[{\textit{Shafranov}(1963)}]{Shafranov_1963}
Shafranov, V.~D., Equilibrium of a toroidal plasma in a magnetic field,
  \textit{J. Nuclear Energy}, \textit{5}, 251--258, 1963.

\bibitem[{\textit{Shen et~al.}(2014)\textit{Shen, Wang, Pan, Miao, Ye, and
  Wang}}]{Shen_etal_2014}
Shen, C., Y.~Wang, Z.~Pan, B.~Miao, P.~Ye, and S.~Wang, Full-halo coronal mass
  ejections: Arrival at the {Earth}, \textit{J. Geophys. Res.: Space Phys.},
  \textit{119}, 5107--5116, 2014.

\bibitem[{\textit{Slavin}(2004)}]{Slavin_2004}
Slavin, J.~A., {Mercury}'s magnetosphere, \textit{Adv. in Space Res.},
  \textit{33}, 1859--1874, 2004.

\bibitem[{\textit{Srivastava et~al.}(2010)\textit{Srivastava, Zaqarashvili,
  Kumar, and Khodachenko}}]{Srivastava_etal_2010}
Srivastava, A.~K., T.~V. Zaqarashvili, P.~Kumar, and M.~L. Khodachenko,
  Observation of kink instability during small {B5.0} solar flare on 2007
  {June} 4, \textit{Astrophys. J.}, \textit{715}, 292--299, 2010.

\bibitem[{\textit{Svedhem et~al.}(2007)}]{Svedhem_etal_2007}
Svedhem, H., et~al., {Venus Express} -- the first {European} mission to
  {Venus}, \textit{Planet. Space Sci.}, \textit{55}, 1636--1652, 2007.

\bibitem[{\textit{Szabo}(1994)}]{Szabo_1994}
Szabo, A., An improved solution to the `{Rankine-Hugoniot}' problem, \textit{J.
  Geophys. Res.: Space Phys.}, \textit{99}, 14,737–--14,746, 1994.

\bibitem[{\textit{Thernisien}(2011)}]{Thernisien_2011}
Thernisien, A., Implementation of the graduated cylindrical shell model for the
  three-dimensional reconstruction of coronal mass ejections,
  \textit{Astrophys. J. Suppl. Ser.}, \textit{194}, 33, 2011.

\bibitem[{\textit{Titov and D\'{e}moulin}(1999)}]{Titov_Demoulin_1999}
Titov, V.~S., and P.~D\'{e}moulin, Basic topology of twisted magnetic
  configurations in solar flares, \textit{Astron. \& Astrophys.}, \textit{351},
  707--720, 1999.

\bibitem[{\textit{{van Ballegooijen} and
  Martens}(1989)}]{vanBallegooijen_Martens_1989}
{van Ballegooijen}, A.~A., and P.~C.~H. Martens, Formation and eruption of
  solar prominences, \textit{Astrophys. J.}, \textit{343}, 971--984, 1989.

\bibitem[{\textit{Vandas and Romashets}(2003)}]{Vandas_Romashets_2003}
Vandas, M., and E.~P. Romashets, A force-free field with constant alpha in an
  oblate cylinder: A generalization of the lundquist solution, \textit{Astron.
  \& Astrophys.}, \textit{398}, 801--807, 2003.

\bibitem[{\textit{Vi{\~{n}}as and Scudder}(1984)}]{Vinas_Scudder_1986}
Vi{\~{n}}as, A.~F., and J.~D. Scudder, Fast and optimal solution to the
  `{Rankine-Hugoniot} problem', \textit{J. Geophys. Res.: Space Phys.},
  \textit{91}, 39--58, 1984.

\bibitem[{\textit{Vourlidas et~al.}(2013)\textit{Vourlidas, Lynch, Howard, and
  Li}}]{Vourlidas_etal_2013}
Vourlidas, A., B.~J. Lynch, R.~A. Howard, and Y.~Li, How many {CMEs} have flux
  ropes? deciphering the signatures of shocks, flux ropes, and prominences in
  coronagraph observations of cmes, \textit{Sol. Phys.}, \textit{284},
  179--201, 2013.

\bibitem[{\textit{Vr\v{s}nak et~al.}(1991)\textit{Vr\v{s}nak, Ruzdjak, and
  Rompolt}}]{Vrsnak_etal_1991}
Vr\v{s}nak, B., V.~Ruzdjak, and B.~Rompolt, Stability of prominences exposing
  helicial-like patterns, \textit{Sol. Phys.}, \textit{136}, 151--167, 1991.

\bibitem[{\textit{Wang et~al.}(2017)\textit{Wang, Liu, Wang, Hu, Shen, Jiang,
  and Zhu}}]{WangW_etal_2017}
Wang, W., R.~Liu, Y.~Wang, Q.~Hu, C.~Shen, C.~Jiang, and C.~Zhu, Buildup of a
  highly twisted magnetic flux rope during a solar eruption, \textit{Nature
  Commun.}, \textit{8}, 1330, 2017.

\bibitem[{\textit{Wang et~al.}(2004)\textit{Wang, Shen, Ye, and
  Wang}}]{Wang_etal_2004b}
Wang, Y., C.~Shen, P.~Ye, and S.~Wang, Deflection of coronal mass ejection in
  the interplanetary medium, \textit{Sol. Phys.}, \textit{222}, 329--343, 2004.

\bibitem[{\textit{Wang et~al.}(2011)\textit{Wang, Chen, Gui, Shen, Ye, and
  Wang}}]{Wang_etal_2011}
Wang, Y., C.~Chen, B.~Gui, C.~Shen, P.~Ye, and S.~Wang, Statistical study of
  coronal mass ejection source locations: Understanding cmes viewed in
  coronagraphs, \textit{J. Geophys. Res.: Space Phys.}, \textit{116}, A04,104,
  doi:{10.1029/2010JA016,101}, 2011.

\bibitem[{\textit{Wang et~al.}(2015)\textit{Wang, Zhou, Shen, Liu, and
  Wang}}]{Wang_etal_2015}
Wang, Y., Z.~Zhou, C.~Shen, R.~Liu, and S.~Wang, Investigating plasma motion of
  magnetic clouds at 1 {AU} through a velocity-modified cylindrical force-free
  flux rope model, \textit{J. Geophys. Res.: Space Phys.}, \textit{120},
  1543--1565, 2015.

\bibitem[{\textit{Wang et~al.}(2016{\natexlab{a}})\textit{Wang, Zhuang, Hu,
  Liu, Shen, and Chi}}]{Wang_etal_2016}
Wang, Y., B.~Zhuang, Q.~Hu, R.~Liu, C.~Shen, and Y.~Chi, On the twists of
  interplanetary magnetic flux ropes observed at 1 {AU}, \textit{J. Geophys.
  Res.: Space Phys.}, \textit{121}, 9316--9339, 2016{\natexlab{a}}.

\bibitem[{\textit{Wang et~al.}(2016{\natexlab{b}})}]{Wang_etal_2016a}
Wang, Y., et~al., On the propagation of a geoeffective coronal mass ejection
  during 15–17 {March} 2015, \textit{J. Geophys. Res.: Space Phys.},
  \textit{121}, 7423–--7434, 2016{\natexlab{b}}.

\bibitem[{\textit{Winslow et~al.}(2015)\textit{Winslow, Lugaz, Philpott,
  Schwadron, Farrugia, Anderson, and Smith}}]{Winslow_etal_2015}
Winslow, R.~M., N.~Lugaz, L.~C. Philpott, N.~A. Schwadron, C.~J. Farrugia,
  B.~J. Anderson, and C.~W. Smith, Interplanetary coronal mass ejections from
  {MESSENGER} orbital observations at {Mercury}, \textit{J. Geophys. Res.:
  Space Phys.}, \textit{120}, 6101--6118, 2015.

\bibitem[{\textit{Winslow et~al.}(2016)\textit{Winslow, Lugaz, Schwadron,
  Farrugia, Yu, Raines, Mays, Galvin, and Zurbuchen}}]{Winslow_etal_2016}
Winslow, R.~M., N.~Lugaz, N.~A. Schwadron, C.~J. Farrugia, W.~Yu, J.~M. Raines,
  M.~L. Mays, A.~B. Galvin, and T.~H. Zurbuchen, Longitudinal conjunction
  between {MESSENGER} and {STEREO A}: Development of {ICME} complexity through
  stream interactions, \textit{J. Geophys. Res.: Space Phys.}, \textit{121},
  6092--6106, 2016.

\bibitem[{\textit{Xiong et~al.}(2006)\textit{Xiong, Zheng, Wang, and
  Wang}}]{Xiong_etal_2006a}
Xiong, M., H.~Zheng, Y.~Wang, and S.~Wang, Magnetohydrodynamic simulation of
  the interaction between interplanetary strong shock and magnetic cloud and
  its consequent geoeffectiveness, \textit{J. Geophys. Res.: Space Phys.},
  \textit{111}, A08,105, 2006.

\bibitem[{\textit{Xiong et~al.}(2007)\textit{Xiong, Zheng, Wu, Wang, and
  Wang}}]{Xiong_etal_2007}
Xiong, M., H.~Zheng, S.~T. Wu, Y.~Wang, and S.~Wang, Magnetohydrodynamic
  simulation of the interaction between two interplanetary magnetic clouds and
  its consequent geoeffectiveness, \textit{J. Geophys. Res.: Space Phys.},
  \textit{112}, A11,103, 2007.

\bibitem[{\textit{Yashiro et~al.}(2004)\textit{Yashiro, Gopalswamy, Michalek,
  {St. Cyr}, Plunkett, Rich, and Howard}}]{Yashiro_etal_2004}
Yashiro, S., N.~Gopalswamy, G.~Michalek, O.~C. {St. Cyr}, S.~P. Plunkett, N.~B.
  Rich, and R.~A. Howard, A catalog of white light coronal mass ejections
  observed by the soho spacecraft, \textit{J. Geophys. Res.: Space Phys.},
  \textit{109}, A07,105, 2004.

\bibitem[{\textit{Zhang et~al.}(2012)\textit{Zhang, Cheng, and
  Ding}}]{Zhang_etal_2012}
Zhang, J., X.~Cheng, and M.-D. Ding, Observation of an evolving magnetic flux
  rope before and during a solar eruption, \textit{Nature Commun.}, \textit{3},
  747, 2012.

\bibitem[{\textit{Zhang et~al.}(2006)}]{ZhangT_etal_2006}
Zhang, T.~L., et~al., Magnetic field investigation of the {Venus} plasma
  environment: Expected new results from {Venus Express}, \textit{Planet. Space
  Sci.}, \textit{54}, 13--14, 2006.

\bibitem[{\textit{Zhuang et~al.}(2017)\textit{Zhuang, Wang, Shen, Liu, Wang,
  Pan, Li, and Liu}}]{ZhuangB_etal_2017}
Zhuang, B., Y.~Wang, C.~Shen, S.~Liu, J.~Wang, Z.~Pan, H.~Li, and R.~Liu,
  Significance of the influence of the {CME} deflection in interplanetary space
  on the {CME} arrival at the {Earth}, \textit{Astrophys. J.}, \textit{845},
  117(12pp), 2017.

\bibitem[{\textit{Zurbuchen and Richardson}(2006)}]{Zurbuchen_Richardson_2006}
Zurbuchen, T.~H., and I.~G. Richardson, In-situ solar wind and magnetic field
  signatures of interplanetary coronal mass ejections, \textit{Space Sci.
  Rev.}, \textit{123}, 31--43, 2006.

\end{thebibliography}

\end{article}
\end{document}